\documentstyle[preprint,aps]{revtex}

\def\beq{\begin{equation}}
\def\eeq{\end{equation}}
\def\beqa{\begin{eqnarray}}
\def\eeqa{\end{eqnarray}}
\def\nn{\nonumber}
\begin{document}
\begin{titlepage}
\begin{flushright}
IC/97/158\\
SISSA REF 129/97/EP
\end{flushright}
\begin{center}
{\Huge Closed string radiation from moving D-branes}
 
\vspace*{.5cm}

{\large  Faheem Hussain$^1$, Roberto Iengo$^2$, Carmen N\'u\~nez$^3$ 
and \\ Claudio A. Scrucca$^{2,4}$}

\vspace*{0.5cm}

$1$ International Centre for Theoretical Physics, Trieste, Italy\\
$^2$ International School for Advanced Studies and INFN, Trieste, Italy\\
$^3$ Instituto de Astronom\'{\i}a y F\'{\i}sica del Espacio (CONICET), 
Buenos Aires, Argentina\\
$^{4}$ Institut de Physique Th\'eorique, Universit\'e de Neuch\^atel,
Switzerland

\vspace*{-.5cm}
 
\end{center} 
 
\begin{abstract}

We compute the amplitude for the radiation of massless NS-NS closed string
states from the interaction of two moving D-branes. We consider
particle-like D-branes with reference to 4-dimensional spacetime, in
toroidal and orbifold compactifications, and we work out the relevant
world sheet propagators within the moving boundary state formalism. We
find no on-shell axion emission. For large inter-brane separation,
we compute the spacetime graviton emission amplitude and estimate 
the average energy radiated, whereas the
spacetime dilaton amplitude is found to vanish in this limit. 
The possibility of emission
of other massless states depends on the nature of the branes and of the
compactification scheme.

\end{abstract}
 
\begin{flushleft}
\end{flushleft}
\end{titlepage}
 
\section{Introduction and Summary}
The non-relativistic dynamics of Dirichlet 
branes \cite{bachas}, plays an essential role in the understanding 
of string theory at scales shorter than the Planck length 
\cite{kp,shenker1,shenker2}. Thus it is important to investigate the physical
properties of the interactions between branes and/or branes and strings.
These physical properties can be generically inferred from scattering
amplitudes. For instance one can look at the interaction of strings with
one brane, by considering a disk-like world sheet with appropriate
boundary conditions and inserting Neveu-Schwarz-Neveu-Schwarz (NS-NS) or
Ramond-Ramond (RR) string vertex operators \cite{disk}. 

Interactions between
two branes can be studied by considering an annulus-like world sheet with
one boundary on each brane. As is well known one finds in particular that
there is no interaction between two identical branes at rest
\cite{pol}. However, non zero amplitudes can occur when one includes in
the system additional string states, by inserting appropriate vertex
operators
in the annulus, and/or when considering branes in relative motion. In
the case of moving branes, the potential falls off like $V^{4}/r^{D-3}$
for small relative velocity $V$ in the maximally supersymmetric case,
whereas it could generically vanish like $V^{2}/r^{D-3}$ for
compactifications breaking some supersymmetry. In addition to the
universal force, there are also additional spin effects proportional to
$V^{4-n}/r^{D-3+n}$ (in the maximally supersymmetric case) which
distinguish between the various components of the 256-dimensional brane
BPS multiplet \cite{harvey,mss}.

In previous works \cite{hin,hins} we have used the boundary state
formalism \cite{polcai} to study, 
in particular, the case of two branes for a Type II
superstring theory compactified on orbifolds, looking at the dependence 
on the relative velocity of the branes' scattering amplitude \cite{hins}, 
which gives important information on the coupling of
the massless fields to the branes. Here we study the interesting
possibility of particle emission from interacting moving branes. Particle
production is quantum mechanically allowed even within the eikonal
approximation where recoil is neglected and the branes are assumed to move
along straight trajectories. This is in particular more and more precise
when dealing with emission of massless states in the limit of small
momentum and large impact parameter. Here we present a systematic study of
this case, namely by computing the amplitudes for the emission of a single
NS-NS massless closed string state (graviton, dilaton or axion) from a
system of two moving D0-branes, in both toroidal and orbifold
compactifications \cite{min} down to four dimensions. 
Actually we have considered
also an interesting case of some Neumann boundary conditions in the
compactified directions, technically describing a D3-brane, but which is
still a particle-like D0-brane with reference to the uncompactified
spacetime.

We find that the amplitude
for axion emission, which gets contributions only from the odd spin
structure in the twisted orbifold sector, vanishes exactly. On the
contrary only the even spin structures
contribute to the emission amplitudes for the dilaton and graviton. We
study these amplitudes in the field theory limit, that is for large
branes' separation. Such a limiting process picks out the massless closed
string states being
exchanged between the branes. In this limit the ``spacetime" dilaton
emission amplitude also vanishes (by ``spacetime" dilaton we refer to
the particle described by the trace of the space-time part of the vertex
operator polarisation tensor). However the graviton emission amplitude
even in this limit is generically different from zero and we present its
computation. Also the amplitude for emission of other massless particles, 
related to the compact directions of the polarisation tensor, can be 
different from zero, depending on the nature of the branes in the 
compactified directions.

We then compare the graviton emission amplitude in the field
theory limit to Feynman diagrams when the two branes exchange a massless
particle, which can either be a scalar, a vector or a graviton, and the
outgoing graviton is emitted by it. We indeed find that the
RR part of the amplitude, in the limit of
large distance from both branes, corresponds to the coupling of the
graviton to the RR
vector being exchanged between them. Similarly we find that in the same
limit the NS-NS part of the amplitude corresponds to the coupling of the
emitted
graviton to the graviton and the scalar being exchanged between the
branes. It is interesting to note that our result for the graviton
emission amplitude corresponds to the sum of various Feynman diagram
contributions, as is usual in string theory, including both 
bremsstrahlung-like
processes where the graviton is directly emitted by the branes,
and processes where it is emitted far from them. 
We also evaluate the average energy $<p>$ radiated by the two branes 
when they pass each other 
at impact parameter $b$ and relative velocity $V$, finding
$<p> \sim g_s^2l_s^2 \frac{V^{1+2n}}{b^3}$, where $g_s,l_s$ are the string
coupling and length, and the integer $n=2,4$ depends on the brane
nature and compactification scheme, that is essentially on the amount
of supersymmetry. If we extrapolate down to the eleven-dimensional Planck
length  $b\sim g_s^{1/3}l_s$
we would find a maximal radiated energy $\sim g_s\frac{V^{1+2n}}{l_s}$, which
one can compare with the estimate of ref. \cite{shenker2}.  
Actually, this extrapolation would be valid for small velocity 
$V < g_s^{2/3}$, see Sec. IVB

Further,
let us note that the graviton amplitude, when the graviton coming out of
the interaction of two branes is off-shell, could be also regarded as a 
first perturbative correction ${\cal O}(1/r^2)$ in the evaluation of 
the gravitational 
field at large distance $r$ from a system of branes. 
In fact, the perturbation 
expansion of the classical solution in terms of tree diagrams, where 
the branes are sources, would give at the second order the graphs of 
Figs. 1,2,3. 
A related issue is the result for the scalar emission amplitude. 
In fact the pattern of scalar couplings is
of crucial importance for the question of whether a non-zero horizon is
produced. From our result, this seems possible for a system of just one
species of D3-branes on an orbifold, since we find these D3-branes to be
uncoupled to any scalars.

The propagators for the world-sheet bosons and fermions in this system are
an important ingredient of the calculation. Since they are not available
in the literature, we outline the computation of these technical tools and
present the resulting expressions which are interesting in their own right
and constitute an important output of our work.

The paper is organised as follows. To make sure that the arguments of
this paper can be followed without being distracted by too many
computations we have kept in the main text only what is strictly
necessary for its understanding and have relegated technical tools  to
appendices. In Sec. II we construct the general
amplitude for particle emission,  carefully separating out the zero mode
contributions and setting up the kinematics. In Sec. III we discuss the
axion emission amplitude and show that it is zero. Sec. IV is devoted
to the
construction of the dilaton and graviton emission amplitudes. In 
Sec. IVA we show that the dilaton amplitude vanishes in the field theory
limit (by dilaton here we mean the massless scalar corresponding to the
trace part of the polarisation tensor, whose traceless part describes the
four dimensional spacetime graviton). In Sec. IVB we present the computation
and the results for the graviton emission amplitude and the estimate
of the energy radiated in this process. 
In Sec. V we consider the field
theory interpretation of the graviton emission amplitude. In Sec. VI we
discuss the case of the emission of other types of massless particles,
corresponding to other components of the ten dimensional polarisation
tensor. Appendix A outlines the construction of the
spacetime part of the boundary state for a moving D-brane. In Appendix B
we use the boosted boundary states to compute the uncompactified part of
the partition functions. This appendix also contains the calculation of
the propagators for bosons and fermions on the cylindrical world sheet
representing the exchange of closed string states between two relatively
moving D-branes.

\section{General Amplitude}

Consider the interaction of two zero branes, moving with velocities 
$V_1$ and $V_2$ respectively, say along the $1$ direction only. We will call
Transverse (T) the other two uncompactified space directions 2,3.  
In the closed string picture the interaction between two branes is viewed
as the exchange of a closed string between two boundary states,
geometrically describing a cylinder. In the present work we use $\tau$ for
the coordinate along the length of the cylinder, $0\leq\tau\leq l$, and
$\sigma$ as the periodic coordinate running from $0$ to $1$. We will
always consider particle like D-branes, that is the time coordinate
satisfies Neumann boundary conditions, whereas the three uncompactified
space coordinates satisfy Dirichlet boundary conditions. 
The emission of a closed string state from these interacting
branes is described by the matrix element of the appropriate vertex
operator sandwiched between the boundary states describing the branes:

\begin{equation}
{\cal A}= \int_0^\infty dl \int_0^l d\tau \sum_s(\pm)
<B,V_1,Y_1| e^{-lH} V(\tau, \sigma) |B,V_2,Y_2>^s\;,
\end{equation}
where the $\sum_s(\pm)$ represents the sum over the spin structures with
the appropriate signs (GSO projections).

The vertex operator for a massless NS-NS state (this 
state represents a massless particle propagating in
4-dimensional uncompactified spacetime, with momentum $p^{\mu},\,\,
\mu=0,1,2,3$ and $p^{\mu}p_{\mu}=0$) is given by 
\begin{equation}
V(z, \bar z) = 
e_{ij} (\partial X^i - {1\over 2}p\cdot \psi \psi^i) 
(\bar \partial X^j + {1\over 2}p \cdot \bar\psi \bar \psi^j) 
e^{i p \cdot X}\;,
\end{equation}
with $z=\sigma+i\tau$ and $\partial=\partial_z$.
For most of this work we consider a polarisation tensor $e_{\mu\nu}$
with components in the uncompactified directions only. In this case in the
vertex operator, only bosonic and fermionic noncompact coordinates
appear. For relative normalisation of $X$ and $\psi$ see Appendix A.
We can take $e_{\mu\nu}$ to have only space
components, which we denote as $i,j$, as this is allowed by the gauge 
invariance of the vertex operator, and moreover $p^i e_{ij}=0$. 
The various cases we will consider are:
 
1) the axion, described by $e_{ij}=b_{ij}$ with $b_{ij}=-b_{ji}$,

2) the dilaton, described by $e_{ij}= \delta_{ij}-\frac{p_i
p_j}{\vec{p}^2}$,

3) the graviton, described by $e_{ij}=h_{ij}$ with $h_{ij}=h_{ji}$ and
$\delta^{ij}h_{ij}=0$.
In Sect. VI we discuss more general polarisation tensors, having also
components in the compactified directions.

As is well known, the D-brane is described by an appropriate boundary 
state \cite{polcai}. We write the boundary state for the moving D-brane as 
\cite{divec}
\begin{equation}
|B,V_2,Y_2>_{s}=\int
\frac{d^{3}q}{(2\pi)^{3}}e^{-i\vec{Y}_2.\vec{q}}
|q^{\mu}_{B}>\otimes |B,sm>^{s}\;,\label{boundary}
\end{equation}
where $|B,sm>^{s}$ is the boundary state constructed from the
Fock space of the bosonic and fermionic string modes $(sm)$ (see Appendix A).
Here $q^{\mu}_{B}$ is the boost of the momentum $(0,q^1,q^2,q^3)$:
\beq
q^\mu_{B} = (\sinh v_2 q^1, \cosh v_2 q^1, \vec q_T)=(\gamma_2 V_2 q^1,
\gamma_2 q^1, \vec q_T)\;,\label{vec}
\eeq
$v_2$ being the rapidity of the brane 2. Similarly $k^{\mu}_{B}$ is defined 
as the boost of the vector $(0,k^1,k^2,k^3)$ with rapidity $v^1$. In our
notation the integration
measure is always defined as $d^3q=dq^1d^2q_{\perp}$ and similarly in the
following $d^3k=dk^1d^2k_{\perp}$. In eq. (\ref{boundary})
we take different from zero only  the spacetime part of the momentum
emitted by the brane (we will be mainly
concerned with the case of large distances, where we can neglect the
configurations having momentum or winding in the compactified directions
different from zero). Separating the zero modes for $\mu=0,1,2,3$
\beq
X^{\mu}(\sigma,\tau)=X^{\mu}_{0m}(\tau)+X^{\mu}_{osc}(\sigma,\tau)\;,
\eeq
where $X^{\mu}_{0m}(\tau)=X^{\mu}_0-iQ^{\mu}\tau$, we can write:
\beqa 
e^{ipX}&=&e^{ipX_{0m}}\circ e^{ipX_{osc}} \;, \nonumber\\
\partial Xe^{ipX}&=&\partial X_{0m}e^{ipX_{0m}}\circ
e^{ipX_{osc}} +e^{ipX_{0m}}\circ \partial X_{osc}e^{ipX_{osc}}\;,
\eeqa
etc. Since in general we have terms of the kind:
\beq
F(X_{0m})e^{ipX_{0m}}\circ G(sm)e^{ipX_{osc}}\;,
\eeq
where $F$ is an expression $(F(X)=1$ or $\partial X^i$ or 
$\bar{\partial}X^j$ or $\partial X^i\bar{\partial} X^j)$ containing 
$X_{0m}$ only and $G(sm)$ contains everything else, that is generically
all the remaining string mode, both bosonic and fermionic,
we can split the computation:
\beqa
&&<B_1,V_1,Y_1|e^{-lH}F(X_{0m})e^{ipX_{0m}}\circ G(sm)
e^{ipX_{osc}}|B_2, V_2, Y_2>\nonumber\\ 
&&\qquad=<F(X_{0m)}e^{ipX_{0m}}><B_1,V_1, sm|e^{-lH(sm)}G(sm)
e^{ipX_{osc}}|B_2, V_2, sm>^s\;.
\eeqa
We have defined:
\beq
<F(X_{0m})e^{ipX_{0m}}>\equiv
\int\frac{d^3 \vec k}{(2\pi)^3}\frac{d^3 \vec q}{(2\pi)^3}
e^{i\vec{Y}_1\vec{k}-i\vec{Y}_2\vec{q}}<k^{\mu}_{B}|e^{-lH(X_{0m})}F(X_{0m})
e^{ipX_{0m}}|q^{\mu}_{B}>\;.
\eeq
To avoid ambiguity let us stress here that $\vec{q}\,\,(\vec{k})$ refers to
the space components of the vector $q^{\mu}(k^{\mu})$ defined in eq.
(\ref{vec}).
As mentioned above, we have four possibilities for $F(X_{0m})$.

Consider first $F=1$, giving
\begin{equation}
<e^{i p \cdot X_{0m}}>_o = \int \frac {d^3 \vec k}{(2 \pi)^3} 
e^{i \vec k \cdot \vec Y_1}
\int \frac {d^3 \vec q}{(2 \pi)^3}
e^{-i \vec q \cdot \vec Y_2}
e^{- \frac l2 k_B^2 + \tau p\cdot q_{B}} <k^\mu_{B}|(p + q_{B})^\mu>\;.
\end{equation}
Notice now that
\begin{eqnarray}
&&<k^\mu_{B}|(p + q_{B})^\mu> = (2 \pi)^4 \delta^{(4)}(p^\mu - k^\mu_{B}
 + q^\mu_{B})) \nonumber \\ 
&&\qquad = \frac {(2 \pi)^4}{\sinh |v_1 - v_2|} 
\delta \left(k^1 - \frac {p^{(2)}}{\sinh (v_1 - v_2)}\right)
\delta \left(q^1 - \frac {p^{(1)}}{\sinh (v_1 - v_2)}\right)
\delta^{(2)}(\vec p_T - \vec k_T + \vec q_T) \;.
\end{eqnarray}
We have used $(V_1 - V_2) \gamma_1 \gamma_2 = \sinh(v_1 - v_2)$ and
defined the boosted energies to be
\begin{equation}
p^{(1,2)} = \gamma_{1,2} (1 - V_{1,2} \cos \theta) p \;,
\end{equation}
where $p=p^0$ and $\cos \theta = \frac {p^1}{p}$. From now we drop the
subscript ``B" on the momenta and use the notation
$q^{\mu}=(q^{0},q^{1},\vec{q}_{T})$, $k^{\mu}=(k^{0},k^{1},\vec{k}_{T})$
where: 
\begin{eqnarray}
&&k^0 = V_1 k^1 \;,\;\; q^0 = V_2 q^1 \;, \nonumber\\
&&k^1 = \frac p{V_1 - V_2} (1 - V_2 \cos \theta) \;,\;\;
q^1 = \frac p{V_1 - V_2} (1 - V_1 \cos \theta) \;, \nonumber\\
&&\vec k_{T} - \vec q_{T} = \vec p_{T} \;.
\end{eqnarray}
Introducing the impact parameter $\vec b = \vec Y_1 - \vec Y_2$
and defining $l^\prime = l - \tau$, finally we get
\begin{equation}
<e^{i p \cdot X_{0m}}> = \frac 1{\sinh |v_1 - v_2|}  
\int \frac {d^2 \vec k_{T}}{(2 \pi)^2} e^{i \vec k \cdot \vec b} 
e^{- \frac {q^2}2 \tau} e^{- \frac {k^2}2 l^\prime} \;.
\end{equation}

The other three possible matrix elements $F(X_{0m})$ are easily
evaluated
since they correspond to further insertions of momentum operators $Q^i$,
and involve additional $k^i$ factors in the integral. They are:
\begin{eqnarray}
&&<\partial X^i_{0m} e^{i p \cdot X_{0m}}> = \frac
1{\sinh |v_1 - v_2|}  
\int \frac {d^2 \vec k_{T}}{(2 \pi)^2} e^{i \vec k \cdot \vec b} 
e^{- \frac {q^2}2 \tau} e^{- \frac {k^2}2 l^\prime} (- \frac 12
k^i) \;, \nonumber \\
&&<\bar \partial X^j_{0m} e^{i p \cdot X_{0m}}> = \frac
1{\sinh |v_1 - v_2|}  
\int \frac {d^2 \vec k_{T}}{(2 \pi)^2} e^{i \vec k \cdot \vec b} 
e^{- \frac {q^2}2 \tau} e^{- \frac {k^2}2 l^\prime} (\frac {1}{2}
k^j) \;, \nonumber\\
&&<\partial X^i_{0m} \bar \partial X^j_{0m} 
e^{i p \cdot X_{0m}}> = 
\frac 1{\sinh |v_1 - v_2|}  
\int \frac {d^2 \vec k_{T}}{(2 \pi)^2} e^{i \vec k \cdot \vec b} 
e^{- \frac {q^2}2 \tau} e^{- \frac {k^2}2 l^\prime}
(- \frac 14 k^i k^j) \;.
\end{eqnarray}

It will prove very convenient to change the integration variables of the
amplitude to $\tau$ and $l^\prime$ which will be interpreted as the 
proper times of the particles mediating the interaction from each of the 
two branes to the vertex:
\begin{equation} 
\int_0^\infty dl \int_0^l d\tau = \int_0^\infty d\tau \int_0^\infty 
dl^\prime \;.
\end{equation}
Notice that $\tau = 0$ corresponds to the
emission from the second brane $|B,V_2,Y_2>$, whereas $l^\prime = 0$, 
corresponds to the emission from the first brane $<B,V_1,Y_1|$;
conversely, $\tau,l^\prime > 0$ corresponds instead to emission far from
both branes.

As for the factor containing the string modes, we write
\beq
<B_1, V_1, sm|e^{-lH(sm)}G(sm)
e^{ipX_{osc}}|B_2, V_2, sm>^s
\equiv
<G(sm)e^{ipX_{osc}}>^s\cdot Z^bZ^{fs}
\label{matrix}
\eeq
where $Z^b$ and $Z^{fs}$ are the partition functions defined by
\beq
Z^b\equiv <B_1,V_1|e^{-lH}|B_2,V_2>_{bosonic\,\, osc}\;,\;\;
Z^{fs}\equiv <B_1,V_1|e^{-lH}|B_2,V_2>_{fermionic\,\, modes}^s \;.
\eeq

For the matrix element of an operator, as defined in eq. (\ref{matrix}),
we distinguish the case of even and odd spin structures. 
In the boundary state formalism, the various spin
structures correspond to the GSO projections and are obtained by the
possibility of inserting the operator $(-1)^F$ in the matrix element of
the boundary states, in the Ramond-Ramond or Neveu-Schwarz-Neveu-Schwarz
case. Since the boundary state is of the form $|B>=e^{i\phi}|0>$, where
$\phi$ is an expression quadratic in the left and right moving fermionic
modes and $|0>$  is a suitable Fock vacuum (see Appendix A), the
insertion of $(-1)^F$ has the effect of changing the sign of $\phi$ in one
of the boundary states. To take it into account, we define
$|B,\eta>=e^{i\eta \phi}|0>$  where $\eta=\pm 1$ (actually, only the
relative sign ${\eta}_1 {\eta}_2$ is relevant, ${\eta}_{1,2}$ referring 
to the $B_{1,2}$ boundary state). The odd spin structure corresponds to 
${\eta}_1 {\eta}_2=-1$ for the RR case. For the even case for any operator
${\cal O}$ we define
\beq <{\cal O}(\sigma,\tau)>^{even}\equiv
\frac{<B_1,V_1,{\eta}_1|e^{-lH}{\cal
O}(\sigma,\tau)|B_2,V_2,{\eta}_2>}
{<B_1,V_1,{\eta}_1|e^{-lH}|B_2,V_2,{\eta}_2>}\;.\label{ave}
\eeq
For the odd case in general there are fermionic zero modes which make the
result zero unless they are soaked up by an equal number of insertions.
Since in our vertex only spacetime fermionic coordinates appear, the
overall result will be zero in the odd case whenever there are zero modes
for the compactified fermionic coordinates. Thus, we restrict the
discussion of the odd spin structure to the case of the $Z_3$ orbifold
when the branes are at the fixed point. In this case, for the twisted sector,
there are no zero modes in the compactified directions \cite{hin,hins}. 
In the boundary state formalism this is seen from Appendix A, because
\beq
<0|e^{i\eta\tilde{b}^{*}b}e^{i\eta b^{*}\tilde{b}}|0>=0
\eeq
where $b=(-i{\gamma}^A+{\gamma}^{A+1})/2$, with $A\geq 2$, is a fermionic
zero mode in a transverse direction,
whereas
\beq
<0|e^{i\eta\tilde{b}^{*}b}bb^{*}e^{i\eta b^{*}\tilde{b}}|0>=1
\eeq
and also different from zero on inserting $b\tilde{b}^{*}$,
$\tilde{b}b^{*}$, and $\tilde{b}\tilde{b}^{*}$. Notice instead that in the
longitudinal direction, $a=(\gamma^0+\gamma^1)/2$,
\beq
\frac{e^{v_2-v_1}}{2}<0|e^{-i\eta e^{2v_1}\tilde{a}a}e^{-i\eta
e^{-2v_2}a^{*}\tilde{a}^{*}}|0>= \sinh (v_2-v_1)\;.
\eeq
Thus in order to have a nonvanishing result, the operator $\cal O$ must
contain $\psi^2_0\psi^3_0$ or $\psi^2_0\tilde{\psi}^3_0$ or
$\tilde{\psi}^2_0\psi^3_0$ or $\tilde{\psi}^2_0\tilde{\psi}^3_0$. In this
case we define (for the RR case with $\eta_1 \eta_2=-1$):
\beq
\ll{\cal O}(\sigma,\tau)\gg^{odd}\equiv \frac{<B_1,V_1,\eta_1|e^{-lH}{\cal
O}(\sigma,\tau)|B_2,V_2,\eta_2>}
{<B_1,V_1,\eta_1|e^{-lH}\psi^2_0\bar\psi^3_0|B_2,V_2,\eta_2>}\;.
\label{aveodd}
\eeq
The fermionic partition function for the odd spin structure is accordingly
defined with the zero modes insertion, as in the denominator of (\ref{aveodd}).
If $\cal O$ does not contain zero modes the result will be zero.

Since $|B>$ can be written as a direct product for pairs of directions
and therefore also the expectation value can be accordingly factorized, we
can evaluate $<{\cal O}>$ by using the Wick theorem for each factor. In
the odd case the factor relative to the $2,3$ directions will be zero if
$\cal O$ does not contain the zero modes. The relevant rules are given in
Appendix B.
In particular
\beq
<G_B(X_{osc})G_F(\psi)e^{ipX_{osc}}>=
<G_B(X_{osc})e^{ipX_{osc}}><G_F(\psi)>
\eeq
and for the expressions containing $X_{osc}$ we have
\beqa
<\partial X^i_{osc}e^{ipX_{osc}}>&=&i<\partial X^i p\cdot X>_{osc}
<e^{ipX_{osc}}>\;,\nonumber\\
<\bar{\partial} X^i_{osc}e^{ipX_{osc}}>&=&i<\bar{\partial} X^i p\cdot 
X>_{osc}<e^{ipX_{osc}}>\;.
\eeqa

We thus get, finally, that the general amplitude can be written as :
\begin{equation}
{\cal A} = \frac 1{\sinh |v_1 - v_2|} \int_0^\infty d\tau 
\int_0^\infty dl^\prime \int \frac {d^2 \vec k_{T}}{(2 \pi)^2} 
e^{i \vec k\cdot \vec b}e^{- \frac {q^2}2 \tau} 
e^{- \frac {k^2}2 l^\prime} <e^{i p \cdot X}>_{osc}{\cal N} \label{amp}
\end{equation}
where
\beq
{\cal N}= Z^b \sum_s(\pm) Z^{fs}{\cal M}^s
\eeq
and
\begin{eqnarray}
\label{ampm}
{\cal M}^s = e_{ij}&& \left\{<\partial X^i \bar \partial X^j>_{osc}
- <\partial X^i p \cdot X>_{osc} <\bar \partial X^j p \cdot X>_{osc} 
\right. \nonumber \\
&&\; +\frac 14 \left(<p \cdot \psi p \cdot \bar \psi>^s 
<\psi^i \bar \psi^j>^s -<p \cdot \psi \psi^i>^s 
<p \cdot \bar \psi \bar \psi^j>^s \right. \nonumber \\ 
&& \qquad \;\left. + <p \cdot \bar \psi \psi^i>^s <p \cdot \psi \bar
\psi^j>^s
\right) \nonumber \\
&&\; + \frac i2 \left(<\partial X^i p \cdot X>_{osc} 
<p \cdot \bar \psi \bar \psi^j>^s
- <\bar \partial X^j p \cdot X>_{osc} <p \cdot \psi \psi^i>^s \right) 
\nonumber \\
&&\; - \frac 12 k^i \left(i<\bar \partial X^j p \cdot X>_{osc} + \frac 12
<p \cdot \bar \psi \bar \psi^j>^s \right) \nonumber \\
&& \left.\; + \frac 12 k^j \left(i<\partial X^i p \cdot X>_{osc} - \frac
12
<p \cdot \psi \psi^i>^{s} \right) - \frac 14 k^i k^j \right\}\;.
\end{eqnarray}
In the case of the odd spin structure, terms not containing $<\psi^2\psi^3>$ 
(or $\bar{\psi}^{2,3}$) at least once are zero, see Appendix B eqs.
(\ref{star}) and (\ref{starodd}).

Notice that the unphysical longitudinal part $b^1$ of the impact parameter
appears in the amplitude only in the irrelevant constant overall phase
$e^{i k^1 b^1}$; one can put $b^1 = 0$ without loss of generality.
In order to get some preliminary physical information from the amplitude, 
we can explicitly carry out the kinematical integration, obtaining
\begin{equation}
\int \frac {d^2 \vec k_{T}}{(2 \pi)^2} e^{i \vec k_{T} \cdot \vec b_T} 
e^{- \frac {q^2}2 \tau} e^{- \frac {k^2}2 l^\prime}=
\frac 1{2\pi l} e^{\frac 1{2l} (\vec p_T \tau - i \vec b_T)^2}
e^{- \frac {p^{(1)2} \tau + p^{(2)2} l^\prime}
{2 \sinh^2 (v_1 - v_2)}}\;.
\end{equation} 
Because of the term $e^{-\frac {\vec b_T^2}{2l}}$, at fixed transverse 
distance $\vec b_T$, world sheets with $l < \vec b_T^2$ are exponentially 
suppressed; in particular, the large distance limit $|\vec b_T| \rightarrow
+\infty$ implies $l \rightarrow +\infty$, and selects the part of
the amplitude where the fields are massless.

In order to complete the computation one has to choose the
compactification scheme. We will consider either toroidal compactification
and D0-branes (that is TypeIIA theory) or $Z_3$ orbifold and either
D0-branes (TypeIIA), or D3-branes (TypeIIB), see our earlier paper
\cite{hins}.

As we will see in the next section the contributions of the even spin
structures to axion emission are zero. The amplitude for the emission of
an axion receives contributions only from the odd spin structure ($RR-$)
sector where one has to insert two transverse zero modes. Of course, in
the case of toroidal compactification from 10 dimensions to 4 dimensions
or, in the case of orbifold compactification, when the D-branes are on a
generic point of the orbifold, the axion production amplitude is trivially
zero, due to the lack of zero mode insertions in the compactified
fermionic coordinates integration. But when the D-branes are on 
the fixed point of a $Z_3$ orbifold there is the twisted sector contribution
where there are no zero modes in the compactified directions. In this case, 
the compactified part of the amplitude turns out to be $``1"$, since the
bosonic part exactly cancels the oscillators of the $RR-$ part and we are 
left with the zero modes' contribution $Z^B Z^{R^-}=4\sinh (v_1-v_2)$.

In the case of the dilaton and the graviton, the situation is different
since the three even spin structures contribute. Now we consider the
D0-brane or also the D3-brane case. Let us first take the D0-brane on a
generic point of the $Z_3$ orbifold or on $T_6$. We will be interested in
the $l\rightarrow \infty$ limit, in which case (see \cite{hins}):
\beqa
&&Z^{NS}_{\pm}\rightarrow e^{2\pi l} \pm 2[\cosh2 (v_1-v_2)+\sum_a \cos 2\pi
z_a]\;,\nonumber\\
&&Z^R_+\rightarrow 16\cosh (v_1-v_2)\prod_a \cos \pi z_a\;.
\eeqa
The case of the D0-brane corresponds to taking all $z_a=0$, whereas for
the D3-branes, with mixed Dirichlet/Neumann boundary conditions in each of
the compactified pairs of coordinates, we take either $z_a=0$ or
$g(z_a=1/3,1/3,-2/3)$
and one has to sum over all possibilities $1+g+g^2$. In this case since
${\cal M}^{NS}_{\pm}=S\pm e^{-2\pi l}T$, where S and T are in general 
functions of $\tau$ and $l^{\prime}$, we finally have  
\beqa
{\cal N}&=&Z^R_{+}{\cal M}^{R+}-Z^{NS}_{+}{\cal M}^{NS+}+Z^{NS}_{-}{\cal
M}^{NS-}
\nonumber\\
&\rightarrow &16\cosh (v_1-v_2)\prod_a \cos\pi z_a{\cal
M}^{R+}-4S[\cosh 2(v_1-v_2)
+\sum_a \cos 2\pi z_a]-2T\;.\label{spinsum}
\eeqa

In the case of a D0-brane on a fixed point of the orbifold there are both
the untwisted sector, where the result is the same as eq. (\ref{spinsum}),
and the twisted sector where one has to make the combination
\beq
{\cal N} = Z^R_{+}{\cal M}^{R+}-Z^{NS}_{+}{\cal M}^{NS+}-
Z^{NS}_{-}{\cal M}^{NS-}\;.
\eeq
Since in this case, for $l\rightarrow\infty$, 
$Z^R_+\rightarrow 2\cosh (v_1-v_2)$, $Z^{NS}_{\pm}\rightarrow 1$, the sum 
over the even spin structures gives
\beq
{\cal N} \rightarrow 2\cosh (v_1-v_2){\cal M}^{R+}-2S \;. 
\eeq

\section{Axion Amplitude}

The axion is described by an antisymmetric and transverse polarisation
tensor satisfying $b^{ij}=-b^{ji}$ and $p_i b^{ij} =  p_j b^{ij} = 0$.
Thus up to a constant $b^{ij}= \frac 12 \epsilon^{ijk}
\frac {p_k}{p}$.

Using the general properties and the definitions given in Appendix B
(remembering that $\epsilon^{ij}\equiv \epsilon^{1ij}$) we have
\begin{eqnarray}
&&<\partial X^i p \cdot X>_{osc} = - <\bar \partial X^i p \cdot
X>_{osc} \;, \nonumber \\
&&<\psi^i \bar \psi^j>^{even} - <\psi^j \bar \psi^i>^{even} =0 \;,\;\;
<p \cdot \psi \psi^j>^{even} - <p \cdot \bar \psi \bar \psi^j>^{even} = 0
\;,\nonumber \\
&&<\psi^i \bar \psi^j>^{odd} - <\psi^j \bar \psi^i>^{odd} = 
 \epsilon^{ij} 
\;,\;\; <p \cdot \psi \psi^j>^{odd}- <p \cdot \bar \psi \bar \psi^j>^{odd} =
 i p_k \epsilon^{kj} \;.   
\end{eqnarray}
Also from the rules of eq. (\ref{star2}) it is easy to see that
\begin{equation}
b_{ij} <p \cdot \bar \psi \psi^i>^{even} <p \cdot \psi \bar \psi^j>^{even} = 0
\;.
\end{equation}
Thus one can see that the whole amplitude (\ref{ampm}), with
$e_{ij}=b_{ij}=-b_{ji}$, is zero for even spin structures.  
Further noticing that 
\begin{equation}
b_{ij} <p \cdot \bar \psi \psi^i>^{odd} <p \cdot \psi \bar \psi^j>^{odd} = 
b_{ij}   p_k \epsilon^{kj}  <p \cdot \bar \psi \psi^i>^{odd} \;,
\end{equation}
the amplitude is found to reduce to
\begin{eqnarray}
{\cal M}^{odd}= \frac 18 b_{ij} &&\left\{
 \epsilon^{ij} <p \cdot \psi p \cdot \bar \psi>^{odd}
\right. \nonumber \\
&&\left. + 4 p_k \epsilon^{kj} \left (<\partial X^i p \cdot X>_{osc}
- \frac 12 <\psi^i p \cdot \bar \psi>^{odd}
- \frac i2 <\psi^i p \cdot \psi >^{odd}
+ \frac i2 k^i \right) \right\}\;.
\end{eqnarray}
By explicit calculation it is seen that the oscillator parts of the last
set of four terms add up to zero, as expected from world-sheet
supersymmetry, and we are left with
\begin{equation}
{\cal M}^{odd} = \frac 18 b_{ij} \left\{
 \epsilon^{ij} <p \cdot \psi p \cdot \bar \psi>^{odd} 
+ 2 i p_k \epsilon^{kj} \left (k^i + 
i <\psi^i p \cdot \bar \psi>^{odd}_o
- <\psi^i p \cdot \psi >^{odd}_o \right) \right\} \;,
\end{equation}
where the subscript $o$ on the fermionic propagator indicates the zero
mode contribution.

Of course, in the case of toroidal compactification from 10 dimensions
to 4 dimensions, this axion production amplitude is trivially zero, due
to the lack of zero mode insertions in the compactified fermionic
coordinates integration. But in the case of a $Z_{3}$ orbifold
compactification, when the 0-branes are on the fixed points of the
orbifold (see Appendix B of \cite{hin}), there are no zero modes in the
compactified directions. We will thus consider this case. 
Using the explicit form of the polarisation tensor, and evaluating the 
zero modes, the amplitude in the $RR-$ sector reduces to
\begin{eqnarray}
{\cal M}^{R-} = &&\frac  18 \cos \theta 
<p \cdot \psi p \cdot \bar \psi>^{R-}_{osc} 
+ \frac i8 \left[\cos \theta \vec p_T\cdot \vec k_T - \sin^2 \theta p
k^1\right]\nonumber \\
&&+\frac i8 p^2 \left[ 2 \cos \theta F^{oR-}_v - (1 + \cos^2 \theta) 
G^{oR-}_v + \sin^2 \theta U^{oR-}_v \right]\;.
\end{eqnarray}

Finally, using the results of Appendix B and remembering the kinematics,
the last expression can be shown to simplify to
\begin{equation}
{\cal M}^{R-} = \frac 18 \cos \theta \left[ 
 <p \cdot \psi p \cdot \bar \psi>^{R-}_{osc} 
+ \frac i2 (k^2 - q^2) \right]
\end{equation}
and using eq. (\ref{sup}) one ends up with
\begin{equation}
{\cal M}^{R-} = \frac i8 \cos \theta \left[ 
-\partial_\tau <p \cdot X(z) p \cdot \bar X (\bar z)>_{osc} 
+ \frac 12 (k^2 - q^2) \right]\;.
\end{equation}

From eqs. (\ref{zbos}) and (\ref{parfunx}) of Appendix B we see that for
the uncompactified parts of the partition functions we have
$$
Z^b Z^{R-}=4 \sinh (v_1 - v_2)
$$
and observing that $\partial_\tau |_{l}=\partial_{\tau} |_{l^\prime} -
\partial_{l^\prime} |_\tau$ the final integrated amplitude, eq.(\ref{amp}), 
for axion emission is seen to be a total derivative
\begin{eqnarray}
\label{Aax}
{\cal A}^{ax}&=& \frac i2 \cos \theta 
\int_0^\infty d\tau \int_0^\infty dl^\prime 
\int \frac {d^2 \vec k_{T}}{(2 \pi)^2} 
e^{i \vec k \cdot \vec b}(\partial_\tau - \partial_{l^\prime}) 
\left\{e^{- \frac {q^2}2 \tau} e^{- \frac {k^2}2 l^\prime} 
<e^{i p \cdot X}>_{osc}\right\} \nonumber \\ &=& 0\;.
\end{eqnarray}
Here, as in the following, possible surface terms at $\tau,l^\prime = 0$ 
have been dropped by making an analytic continuation from $p^2 < 0 $ 
of formula (\ref{exp2}) for $<e^{i p \cdot X}>_{osc}$.

Thus, finally, we find that there is no on-shell axion emission during
the interaction of two moving branes, even in the case of the $Z_{3}$
orbifold compactification. This result is not in contradiction with our
previous work \cite{hin}. There we computed the amplitude for axion
production due to the interaction of an incoming graviton with two
parallel branes at rest. Indeed we found no pole in the axion-graviton
momentum transfer squared and thus there is no on-shell axion coming out 
of the two brane system.

\section{Dilaton and Graviton Amplitudes}

The graviton is described by a symmetric, transverse and traceless 
polarisation tensor, satisfying $h^{ij}=h^{ji}$, 
$p_i h^{ij} = 0$ and $h^i_i = 0$. Consequently, there
are two physical transverse polarisations.
The dilaton, instead, can be thought of as the trace part of the graviton
and is 
described by a symmetric and transverse polarisation tensor, 
satisfying $h^{ij}=h^{ji}$ and $p_i h^{ij} = 0$,
which can be taken to be $h^{ij} =
\delta^{ij}-
\frac {p^i p^j}{p^2}$. 

In these cases one can verify that, due to the symmetry of the polarisation 
tensor, the amplitudes are non-vanishing in the even spin structure sectors 
only.

It will prove of great help in this case to integrate by parts the 
two-derivative bosonic term; by using 
$\bar \partial = \frac i2 \partial_\tau |_{l}=
\frac i2 (\partial_{\tau} |_{l^\prime} - \partial_{l^\prime} |_\tau)$, 
since $\bar \partial$ acts on a function of $z - \bar z = 2i\tau$, and 
observing that the partition function behaves like a constant with respect 
to the latter derivative since it depends only on $l = \tau + l^\prime$,
one gets
\begin{eqnarray}
&&\int_0^\infty d\tau \int_0^\infty dl^\prime e^{- \frac {q^2}2 \tau} 
e^{- \frac {k^2}2 l^\prime} <e^{i p \cdot X}>_{osc} 
h_{ij}<\partial X^i(z) \bar \partial \bar X^j(\bar z)>_{osc} \nonumber \\
&& \qquad = - \frac i2 \int_0^\infty d\tau \int_0^\infty dl^\prime
 h_{ij}<\partial X^i(z) \bar X^j(\bar z)>_{osc}
(\partial_{\tau} - \partial_{l^\prime}) \left\{
e^{- \frac {q^2}2 \tau} e^{- \frac {k^2}2 l^\prime} 
<e^{i p \cdot X}>_{osc} \right\} \nonumber \\
&& \qquad = - \int_0^\infty d\tau \int_0^\infty dl^\prime
e^{- \frac {q^2}2 \tau} e^{- \frac {k^2}2 l^\prime} 
<e^{i p \cdot X}>_{osc} 
h_{ij} <\partial X^i(z) \bar X^j(\bar z)>_{osc} \nonumber \\
&& \qquad \qquad \qquad \qquad \qquad \times 
\left\{<p \cdot \partial X(z) p \cdot \bar X(\bar z)>_{osc}
+ \frac i4 (k^2 - q^2) \right\}\;.
\end{eqnarray}
Taking into account also the symmetry of $h^{ij}$ for both the 
graviton and the dilaton and the property (see appendix B)
\begin{equation}
<\partial X^i p \cdot X>_{osc} = - <\bar \partial X^i p \cdot X>_{osc}\;,
\end{equation}
the amplitude ${\cal M}^{s}$ in (\ref{ampm}) can be taken to be (writing
$X_{osc}(z,\bar z) = X_{osc}(z) + \bar X_{osc}(\bar z)$) 
\begin{eqnarray}
{\cal M}^{s} = h_{ij}&& \left\{-<\partial X^i \bar X^j>_{osc}
<p \cdot \partial X p \cdot \bar X>_{osc} \right. \nonumber \\
&&\;\; + <\partial X^i p \cdot (X + \bar X)>_{osc} 
<\partial X^j p \cdot (X + \bar X)>_{osc} \nonumber \\
&&\; +\frac 14 \left(<p \cdot \psi p \cdot \bar \psi>^s 
<\psi^i \bar \psi^j>^s -<p \cdot \psi \psi^i>^s 
<p \cdot \bar \psi \bar \psi^j>^s \right. \nonumber \\ 
&& \qquad \;\left. + <p \cdot \bar \psi \psi^i>^s 
<p \cdot \psi \bar \psi^j>^s \right) \nonumber \\
&&\; + \frac 12 \left(i<\partial X^i p \cdot (X + \bar X)>_{osc} 
+ \frac 12 k^i \right) \left(<p \cdot \psi \psi^j>^s 
+ <p \cdot \bar \psi \bar \psi^j>^s \right) \nonumber \\
&& \left.\; + i k^i <\partial X^j p \cdot (X + \bar X)>_{osc} 
- \frac i4 (k^2 - q^2) <\partial X^i \bar X^j >_{osc}
- \frac 14 k^i k^j \right\}\;. 
\end{eqnarray}

We will focus on the large distance limit $l \rightarrow +\infty$, 
in which only the massless modes will contribute and we expect the
low energy effective field theory to reproduce all the results.
Since $l=\tau + l^\prime$, in this limit at least one among $\tau$ and 
$l^\prime$ is large and thus a massless state is propagating between the
two branes, which are far away from each other. If $l^\prime \rightarrow
\infty$ and $\tau$ is finite, the particle is emitted near the second brane;
if $\tau \rightarrow \infty$ and $l^\prime$ is finite, it is emitted near
the first brane. If both $\tau,l^\prime \rightarrow \infty$, the particle is 
emitted far from both branes.

In the large distance limit $l\rightarrow \infty$ the bosonic exponential
reduces to (see eqs. (\ref{exp}) and (\ref{exp2}))
\begin{equation}
<e^{i p \cdot X}>_{osc} = 
\left[1 - e^{-4\pi \tau} \right]^{-\frac {p^{(2)2}}{2\pi}}
\left[1 - e^{-4\pi l^\prime} \right]^{-\frac
{p^{(1)2}}{2\pi}}\;.\label{explim}
\end{equation}
After having evaluated the limiting forms of ${\cal N}$, one has to 
integrate in eq. (\ref{amp}) over the proper times $\tau$ and $l^\prime$,
\beq
 \int_0^\infty d\tau \int_0^\infty dl^\prime e^{- \frac {q^2}2 \tau} 
e^{- \frac {k^2}2 l^\prime} <e^{i p \cdot X}>_{osc}{\cal N}\;.
\label{red}
\eeq
These last integrations will eventually produce factors like $1/q^2$ or
$1/k^2$ or both, corresponding to the denominator of the propagators of
the massless particles emitted by the branes.

\subsection{Dilaton}

Using the explicit form for the polarisation tensor and recalling the 
notation defined in the Appendix B, the amplitude is found to be
\begin{eqnarray}
{\cal M}^{s} = &&\frac {p^2}4 \left\{
\sin^2 \theta \left[(K_v - K)^2 - (F_v^s - F^s)^2 - L_v^2 + G_v^{s2}  
-(U_v^s - W_v)^2 \right. \right. \nonumber \\
&& \qquad \qquad \quad \left. - 2 (U_v^s - W_v) 
[L_v - \cos \theta (K_v - K)] \right] \nonumber \\
&& \left. \qquad + 4(K K_v - F^s F_v^s) - 4 \cos \theta (K L_v - F^s G_v^s)
\right\} \nonumber \\
&& + \frac 18 (k^2 - q^2) [\sin^2 \theta K_v + (1 + \cos^2 \theta) K]
\nonumber \\
&& + \frac p2 h_{i1}k^i[L_v - \cos \theta (K_v - K) + (U_v^s - W_v)]
- \frac 14 h_{ij}k^i k^j \;. \label{dilaton}
\end{eqnarray}
For the three even spin structures, 
this expression can be further simplified using the results of Appendix B in 
the $l \rightarrow +\infty$ limit. The non exponential
terms $-\frac {(v_1 - v_2)}{2\pi l}$, present in both $U_v^s$ and $W_v$,
cancel in all the three even spin structures.  

By using the kinematics and the results of Appendix B, the ${\cal
M}^{R+}$ amplitude for $l \rightarrow \infty$ reduces to
\begin{eqnarray}
\label{diR}
{\cal M}^{R+}=&&-\frac 14 h_{ij}k^i k^j \nonumber \\
&& -\left[p^{(2)2} + V_2 \gamma_2 p^{(2)} h_{i1}k^i 
+ \frac 14 (k^2 - q^2)(1 + V_2^2 \gamma_2^2 \sin^2 \theta) \right]
f(\tau) \nonumber \\
&& -\left[p^{(1)2} - V_1 \gamma_1 p^{(1)} h_{i1}k^i
- \frac 14 (k^2 - q^2)(1 + V_1^2 \gamma_1^2 \sin^2 \theta) \right]
f(l^{\prime}) \nonumber \\
&& + \frac 12 \tanh (v_1 - v_2) \cos \theta \left\{- \frac 14 (k^2 - q^2)
+p^{(2)2} f(\tau)-p^{(1)2} f(l^{\prime}) \right\}\;.\label{R+} 
\end{eqnarray}
We define here and in the following
\beq
f(\tau)=\frac{e^{-4\pi \tau}}{1-e^{-4\pi\tau}} \;,\;\;
f(l^{\prime})=\frac{e^{-4\pi l^{\prime}}}{1-e^{-4\pi l^{\prime}}}\;.
\eeq
The last term in ${\cal M}^{R+}$ is easily seen to be a total derivative; 
in fact by inserting into (\ref{red}) both eq. (\ref{explim}) and these last 
terms in eq. (\ref{R+}) we get 
\begin{eqnarray}
&&\int_0^\infty d\tau \int_0^\infty dl^\prime e^{- \frac {q^2}2 \tau} 
e^{- \frac {k^2}2 l^\prime} 
\left[1 - e^{-4\pi \tau} \right]^{-\frac {p^{(2)2}}{2\pi}}
\left[1 - e^{-4\pi l^\prime} \right]^{-\frac {p^{(1)2}}{2\pi}}
\nonumber \\ && \qquad \qquad \qquad \times
\left\{- \frac 14 (k^2 - q^2)
+p^{(2)2} \frac {e^{-4\pi \tau}}{1 - e^{-4\pi \tau}}
-p^{(1)2} \frac {e^{-4\pi l^\prime}}{1 - e^{-4\pi l^\prime}} \right\}
\nonumber \\ && \qquad = -\frac 12
\int_0^\infty d\tau \int_0^\infty dl^\prime \; (\partial_\tau -
\partial_{l^\prime}) \left\{e^{- \frac {q^2}2 \tau} 
e^{- \frac {k^2}2 l^\prime} 
\left[1 - e^{-4\pi \tau} \right]^{-\frac {p^{(2)2}}{2\pi}}
\left[1 - e^{-4\pi l^\prime} \right]^{-\frac {p^{(1)2}}{2\pi}}
\right\} \nonumber \\
&& \qquad =0 \;.
\end{eqnarray}
For later use, notice that this kind of integration by parts implies the
following equivalence relations in the amplitude ${\cal M}^{s}$ (see the
remark made after eq. (\ref{Aax}))
\begin{equation}
\frac {e^{-4\pi \tau}}{1 - e^{-4\pi \tau}} \sim
- \frac 14 \frac {q^2}{p^{(2)2}} \;,\;\;
\frac {e^{-4\pi l^\prime}}{1 - e^{-4\pi l^\prime}} \sim 
- \frac 14 \frac {k^2}{p^{(1)2}}\;.
\label{equiv}
\end{equation}

From kinematics one finds the relations
\begin{eqnarray}
\label{kin}
&&p^{(2)2} + V_2 \gamma_2 p^{(2)2} h_{i1}k^i
= - \frac 12 \cos \theta (k^2 - q^2) V_2 \gamma_2 \frac {p^{(2)}}{p}
+ \frac {k^0}{p} p^{(2)2} \;,\nonumber\\
&&p^{(1)2} - V_1 \gamma_1 p^{(1)2} h_{i1}k^i 
= \frac 12 \cos \theta (k^2 - q^2) V_1 \gamma_1 \frac {p^{(1)}}{p}
- \frac {q^0}{p} p^{(1)2} \;,\nonumber\\ 
&&h_{ij} k^i k^j = -\frac 1{4 p^2} (k^2 - q^2)^2 
+ \frac {k^0}{p} q^2 - \frac {q^0}{p} k^2\;.
\end{eqnarray}
 By using these relations and again the equivalence relations
(\ref{equiv}) one can see that also the remaining terms in (\ref{R+})
cancel, and thus there is no contribution from the RR sector in
$l\rightarrow \infty$ limit.

In the NSNS$\pm$ spin structures the amplitude is found to be
\begin{eqnarray}
\label{diNS}
{\cal M}^{NS\pm}=&&-\frac 14 h_{ij}k^i k^j \nonumber \\
&& -\left[p^{(2)2} + V_2 \gamma_2 p^{(2)} h_{i1}k^i 
+ \frac 14 (k^2 - q^2)(1 + V_2^2 \gamma_2^2 \sin^2 \theta)
\right]f(\tau) \nonumber \\
&& -\left[p^{(1)2} - V_1 \gamma_1 p^{(1)} h_{i1}k^i
- \frac 14 (k^2 - q^2)(1 + V_1^2 \gamma_1^2 \sin^2 \theta) \right]
f(l^{\prime}) \nonumber \\
&& \mp e^{-2\pi l} \left\{p^2 \sin^2 \theta \sinh^2 (v_1 - v_2) 
+ p^{(1)2} + p^{(2)2}  - p h_{i1}k^i \sinh 2(v_1 - v_2) 
\right. \nonumber \\
&& \qquad \quad \;\; \left. -2\cos \theta \sinh 2(v_{1}-v_{2})p^{(2)2}f(\tau) +
2\cos \theta \sinh 2(v_{1}-v_{2})p^{(1)2}f(l^{\prime}) \right\}\;.
\end{eqnarray}
Here we have used the relation
\beq
\frac{1}{1-e^{-4\pi \tau}}\cdot \frac{1}{1-e^{-4\pi l^{\prime}}}=1+
f(\tau) +f(l^{\prime})\;,
\eeq
up to terms ${\cal O}(e^{-4\pi l})$ which we neglect in the large distance 
limit.

The first three rows in equation (\ref{diNS}) are identical to the first
three rows of eq. (\ref{R+}) and thus they also cancel in the integration of
eq. (\ref{red}). Moreover:
\beqa
&&p^2 \sin^2 \theta \sinh^2 (v_1 - v_2) 
+ p^{(1)2} + p^{(2)2}  - p h_{i1}k^i \sinh 2(v_1 - v_2) \nonumber \\
&& \qquad =\frac{1}{2} \cos \theta \sinh 2(v_{1}-v_{2})(k^{2}-q^{2})\;.
\eeqa
Thus modulo terms which cancel in the integration of eq. (\ref{red}), we
are left with ${\cal M}^{NS\pm}~=~\pm~e^{-2\pi l}T$ (following the notation of
eq. (\ref{spinsum})) where
\beq
T= - \cos \theta \sinh 2(v_{1}-v_{2})
\left[\frac{1}{2}(k^{2}-q^{2})-2p^{(2)2}f(\tau)+2p^{(1)2}f(l^{\prime})\right]
\;,
\eeq
which is also seen to be zero using the equivalence relations
(\ref{equiv}). In
conclusion, there is no dilaton emission from interacting moving branes far 
from each other.
 
\subsection{Graviton}

Using the properties of the polarisation tensor for the graviton,
the general amplitude is found to be
\begin{eqnarray}
{\cal M}^{s} = &&\frac {p^2}4 h_{11} \left\{(K_v^{2} - K^{2} - L_v^{2})-
(F_v^{s2} - F^{s2} - G_v^{s2}) -(U_v^s - W_v)^2 \right. \nonumber \\
&& \qquad \quad \; \left. - 2 (U_v^s - W_v) 
[L_v - \cos \theta (K_v - K)] \right\} \nonumber \\
&& + \frac 18 (k^2 - q^2) h_{11} [K_v - K]
\nonumber \\
&& + \frac p2 h_{i1}k^i[L_v - \cos \theta (K_v - K) + (U_v^s - W_v)]
- \frac 14 h_{ij}k^i k^j \;. \label{gr}
\end{eqnarray}

For the three even spin structures, this expression
can be further simplified using the results of Appendix B in the $l 
\rightarrow +\infty$ limit. Just as for the dilaton the non exponential
terms $-\frac {(v_1 - v_2)}{2\pi l}$, present in both $U_v^s$ and $W_v$,
cancel in all the three even spin structures. 
 
By using the kinematics, the amplitude for 
$l \rightarrow \infty$ reduces to
\begin{eqnarray}
\label{grR}
{\cal M}^{R+}=&&-\frac 14 h_{ij}k^i k^j 
 -\left[V_2 \gamma_2 p^{(2)} h_{i1}k^i 
+ \frac 14 (k^2 - q^2) V_2^2 \gamma_2^2 h_{11} \right]f(\tau) \nonumber \\
&& +\left[V_1 \gamma_1 p^{(1)} h_{i1}k^i
+ \frac 14 (k^2 - q^2) V_1^2 \gamma_1^2 h_{11} \right]f(l^{\prime}) 
\nonumber \\
&& + \frac p2 \tanh (v_1 - v_2) \left\{\frac 12 h_{i1} k^i
+V_2 \gamma_2 p^{(2)} h_{11}f(\tau) 
-V_1 \gamma_1 p^{(1)} h_{11}f(l^{\prime})  \right\}\;. 
\end{eqnarray}
For the NS$\pm$ sectors, in the $l\rightarrow \infty$ limit, we get:
\begin{eqnarray}
\label{grNS}
{\cal M}^{NS\pm}=&&-\frac 14 h_{ij}k^i k^j 
 -\left[V_2 \gamma_2 p^{(2)} h_{i1}k^i 
+ \frac 14 (k^2 - q^2) V_2^2 \gamma_2^2 h_{11} \right]f(\tau) \nonumber \\
&& +\left[V_1 \gamma_1 p^{(1)} h_{i1}k^i
+ \frac 14 (k^2 - q^2) V_1^2 \gamma_1^2 h_{11} \right]f(l^{\prime}) \nonumber \\
&& \mp e^{-2\pi l} \left\{p^2 h_{11} \sinh^2 (v_1 - v_2) 
- p h_{i1}k^i \sinh 2(v_1 - v_2) \right. \nonumber \\
&& \qquad \quad \;\; - 2h_{11} V_2 \gamma_2 \sinh 2(v_1 - v_2) p p^{(2)}  
f(\tau) \nonumber \\
&& \qquad \quad \; \left.+2 h_{11} V_1 \gamma_1 \sinh 2(v_1 - v_2) p p^{(1)} 
f(l^{\prime}) \right\}\;.
\end{eqnarray}

The graviton emission amplitude is generically different from zero. 
We can always use the relations (\ref{equiv}) to reduce the final integration
over the two proper times $\tau$ and $l^\prime$ to the expression
(see eq. (\ref{explim}))
\beq
\int_0^{\infty} d\tau \int_0^{\infty}dl^{\prime}
e^{-\frac{q^2}{2}\tau}e^{-\frac{k^2}{2}l^{\prime}}<e^{ip\cdot X}>_{osc}
=I_1I_2
\eeq
where
\beq
I_1=-\frac{1}{4\pi}\frac{\Gamma [\frac{k^2}{8\pi}]\Gamma [-\frac{p^{(1)2}}
{2\pi}+1]}{\Gamma [\frac{k^2}{8\pi}-\frac{p^{(1)2}}{2\pi}+1]}
\rightarrow -\frac 2{k^2}\;,\;\;
I_2=-\frac{1}{4\pi}\frac{\Gamma [\frac{q^2}{8\pi}]\Gamma [-\frac{p^{(2)2}}
{2\pi}+1]}{\Gamma [\frac{q^2}{8\pi}-\frac{p^{(2)2}}{2\pi}+1]}
\rightarrow -\frac 2{q^2}\;.
\eeq
We have indicated the limiting expressions of $I_{1,2}$ for the relevant 
case where the energy $p$ of the emitted graviton is much smaller than the 
string scale. 

Finally the amplitude at fixed impact parameter $\vec b_T$ can be written as
\beqa
\label{stramplgr}
&&{\cal A} = \frac 4{\sinh |v_1 - v_2|}
\int \frac {d^2 \vec k_T}{(2 \pi)^2} e^{i \vec k_T \cdot \vec b_T} 
\left(\frac {B}{q^2 k^2} + \frac {C}{q^2} + \frac {D}{k^2} \right)
\eeqa
remembering that 
\beq
k^2 = \vec k_T^2 + \frac {p^{(2)2}}{\sinh^2 (v_1 - v_2)} \;,\;\;
q^2 = (\vec k_T - \vec p_T)^2 + \frac {p^{(1)2}}{\sinh^2 (v_1 - v_2)}\;.
\eeq
For the untwisted sector one has:
\beqa
B=&&-  h_{ij}k^i k^j \left(4 \cosh (v_1 -  v_2) \prod_a \cos \pi z_a
-[\cosh 2(v_1 - v_2) + \sum_a \cos 2 \pi z_a] \right) \nn \\
&&+ p h_{i1} k^i \left(4 \sinh (v_1 - v_2) \prod_a \cos \pi z_a
-2 \sinh 2(v_1 - v_2) \right) \nn \\
&&+ 2 p^2 h_{11} \sinh^2 (v_1 - v_2)\nn \;, \\
C = &&-  V_1 \gamma_1 \frac {h_{i1}k^i}{p^{(1)}}
\left(4 \cosh (v_1 -  v_2) \prod_a \cos \pi z_a
-[\cosh 2(v_1 - v_2) + \sum_a \cos 2 \pi z_a] \right)\nn \\
&&+  V_1 \gamma_1 \frac {p h_{11}}{p^{(1)}}
\left(2 \sinh (v_1 - v_2) \prod_a \cos \pi z_a -  \sinh 2(v_1 - v_2) \right)
\nn \;, \\
D = &&  V_2 \gamma_2 \frac {h_{i1}k^i}{p^{(2)}}
\left(4 \cosh (v_1 -  v_2) \prod_a \cos \pi z_a
-[\cosh 2(v_1 - v_2) + \sum_a \cos 2 \pi z_a] \right)\nn \\
&&-  V_2 \gamma_2 \frac {p h_{11}}{p^{(2)}}
\left(2 \sinh (v_1 - v_2) \prod_a \cos \pi z_a - \sinh 2(v_1 - v_2) \right)\;.
\eeqa
In $C$ and $D$ we neglected terms of order $k^2$ and $q^2$.

For the twisted sector we have:
\beqa
B=&&- \frac 12 h_{ij}k^i k^j \left( \cosh (v_1 -  v_2)-1 \right) 
+ \frac{p}{2} h_{i1} k^i \sinh (v_1-v_2) \nn \;,\\
C = &&- \frac 12 V_1 \gamma_1 \frac {h_{i1}k^i}{p^{(1)}}
\left( \cosh (v_1 -  v_2)- 1\right)+ \frac 14 V_1 \gamma_1 
\frac {ph_{11}}{p^{(1)}} \sinh (v_1 - v_2) \nn \;,\\
D = && \frac 12 V_2 \gamma_2 \frac {h_{i1}k^i}{p^{(2)}}
\left(\cosh (v_1 -  v_2)- 1\right)- \frac 14 V_2 \gamma_2 
\frac {p h_{11}}{p^{(2)}} \sinh (v_1 - v_2)\;.
\eeqa
This result, which exhibits the graviton emission amplitude from interacting 
moving branes at large distance, comes from a single string diagram.
It receives contributions from three distinct physical processes, namely the 
bremsstrahlung-like case where the graviton is emitted directly from one of 
the branes (C and D terms with a single pole in either $q^2$ or $k^2$),
and the case where it is emitted far from both branes as depicted in the 
figures of next section (B term with poles in both $q^2$ and $k^2$). 

The result simplifies for $\theta=0$, where 
\beq
B=-  h_{ij}k^i k^j \left(4 \cosh (v_1 -  v_2) \prod_a \cos \pi z_a
-[\cosh 2(v_1 - v_2) + \sum_a \cos 2 \pi z_a] \right)
\eeq
for the untwisted sector,
\beq
B=- \frac 12 h_{ij}k^i k^j \left( \cosh (v_1 -  v_2)-1 \right)
\eeq
for the twisted sector and
\beq
C=D=0\;.
\eeq
In this case, we find that the bremsstrahlung terms cancel.

We discuss now the result for small relative velocity $V = \tanh (v_1 - v_2)$.
In the case of the 0-brane, all the $z_a$'s can be set to zero (untwisted
sector) and one finds
\beqa
B \simeq && \frac {V^4}2 h_{ij}k^i k^j - 2 V^3 p h_{i1} k^i 
+ 2 V^2 p^2 h_{11} \nn \;,\\
C \simeq && \frac {V_1 \gamma_1}{p^{(1)}} \left(\frac {V^4}2 h_{i1}k^i 
- V^3 p h_{11} \right) \nn \;,\\
D \simeq && - \frac {V_2 \gamma_2}{p^{(2)}} 
\left(\frac {V^4}2 h_{i1}k^i - V^3 p h_{11} \right)\;. 
\eeqa
Observe that for small velocities $p^{(1,2)} \rightarrow p$.

In the orbifold twisted sector we have 
\beqa
B \simeq && -\frac 14 V^2 h_{ij}k^i k^j + \frac 12 V p h_{i1} k^i  \nn \;,\\
C \simeq && \frac {V_1 \gamma_1}{p^{(1)}} \left(\frac 14 V^2 h_{i1}k^i - 
\frac 14 V p h_{11} \right) \nn \;,\\
D \simeq && - \frac {V_2 \gamma_2}{p^{(2)}} 
\left(\frac 14 V^2 h_{i1}k^i - \frac 14 V p h_{11} \right)\;. 
\eeqa
In the case of the 3-brane of the $Z_3$ orbifold, averaging over the orbifold 
relative twists $z_a$ \cite{hins}, one gets 
$<\prod_a \cos \pi z_a > = \frac 14$, $<\sum_a \cos 2 \pi z_a> =
0$. Thus for $V \rightarrow 0$ one finds
\beqa
B \simeq && \frac 32 V^2 h_{ij}k^i k^j - 3 V p h_{i1} k^i 
+ 2 V^2 p^2 h_{11} \nn \;,\\
C \simeq && \frac {V_1 \gamma_1}{p^{(1)}} \left(\frac 32 V^2 h_{i1}k^i - 
\frac 32 V p h_{11} \right) \nn \;,\\
D \simeq && - \frac {V_2 \gamma_2}{p^{(2)}} 
\left(\frac 32 V^2 h_{i1}k^i - \frac 32 V p h_{11} \right) \;.
\eeqa

Finally, by looking at the nearest singularity in $k^2_T$ in the integrand of
eq. (\ref{stramplgr}) one can estimate that the amplitude is maximal for
$\theta=0$ and that for small $V$
\beq
\label{ordine}
{\cal A} \lesssim V^{n-1} g_sl_s f(b_T\cdot p/V) e^{- b_T \cdot p/V}\;.
\eeq
where $f$ is some mildly varying function, $n=2$ for the D3-brane
on the orbifold while generically $n=4$, and we have inserted 
explicitly the appropriate factors of the
string coupling constant $g_s$ and the string length $l_s$
(assuming a compactification radius of the order of $l_s$).
The cross-section for radiating a particle is
\beq
\sigma = \int d^2 \vec b_T \int \frac{d^3p}{p} |{\cal A}|^2 \;.
\eeq
The probability that two interacting branes at impact parameter $\vec b_T$
and with relative velocity $V$ would radiate a particle is thus
\beq
\label{distr}
dP(p,\vec b_T,V)= |{\cal A}|^2 \frac{d^3p}{p} \;. 
\eeq
Thus the amount of radiated energy is
\beq
\label{stima}
<p> \sim g_s^2 l_s^2 \frac{V^{3+2(n-1)}}{b_T^3} \;.
\eeq
If we extrapolate down to the eleven-dimensional Planck length 
$b_T \sim l_{11} = g_s^{1/3}l_s$ we get a maximal radiated energy 
\beq
<p> \sim g_s \frac{V^{1+2n}}{l_s} 
\eeq
which could be compared with the estimate of ref. \cite{shenker2}
(the power of $V$ in our expression is due to the amount of 
supersymmetry cancellation). Actually, there are dynamical effects,
that we have so far disregarded in our approximations, which could
invalidate the above extrapolation; but for small velocity
they are in fact negligible. First, the dynamics would change
due to open string pair creation \cite{bachas} (a process which is encoded
in the poles of the partition function eq. (B3)) but this effect is 
suppressed even at $b_T \sim l_{11}$ for $V < g_s^{2/3}$ \cite{shenker2}.
Further, the factor $(V/b_T)^3$ in (\ref{stima}) is due the "kinematical"
cutoff on the maximal emitted energy $p_{max} \sim V/b_T$, encoded
in the expression for $dP(p,\vec b_T,V)$ through the exponential factor
of eq. (\ref{ordine}); but there is also the string cutoff $1/l_s$ which 
does not appear here, due to our neglect of the exchange of massive 
string states. However the string cutoff is larger than the kinematical one, 
at $b_T \sim l_{11}$, if $V <g_s^{2/3} < g_s^{1/3}$. Note also that 
in this case $p_{max}$ is much smaller than the brane momentum 
$M_{br}V=V/(g_sl_s)$ and thus the eikonal approximation holds.   

\section{Field Theory Interpretation}

Let us consider the terms in the graviton emission amplitude which have
simultaneously a pole in $k^{2}$ and $q^{2}$. They correspond to
diagrams in which the two branes exchange a massless particle, which can
either be a scalar, a vector or a graviton, and the outgoing graviton is
emitted by it (see Figs. 1,2,3). 

\medskip
\input epsf
\epsfsize=10cm
\centerline{\epsffile{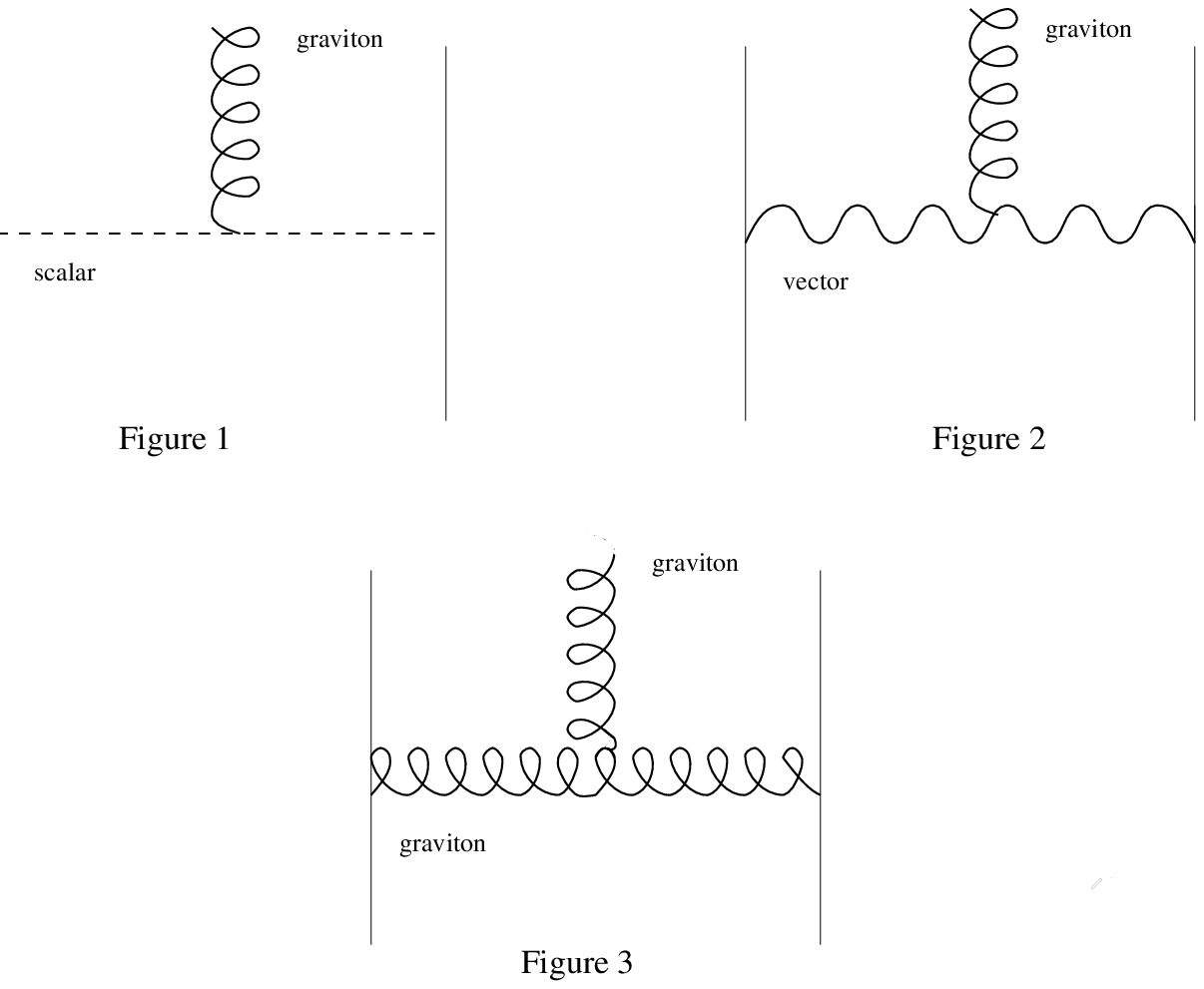}}

The kinematics for Feynman diagrams is the following. The
incoming momenta of the two branes are
\begin{equation}
B_1^\mu = (\gamma_1, V_1 \gamma_1, 0) \;,\;\;
B_2^\mu = (\gamma_2, V_2 \gamma_2, 0)\;.
\end{equation}
Observe that 
\beq
k \cdot B_1 = q \cdot B_2 = 0\;.\label{cons}
\end{equation}
The current, $J^{\mu}$, and energy-momentum tensor, $T^{\mu\nu}$, of the
branes are (neglecting corrections due to the small momentum transfer)
\begin{eqnarray}
&&J_1^\mu = B_1^\mu \;,\;\; J_2^\mu = B_2^\mu \;,\nonumber\\
&&T_1^{\mu\nu} = B_1^\mu B_1^\nu \;,\;\; T_2^{\mu\nu} = 
B_2^\mu B_2^\nu \;,
\end{eqnarray}
and their conservation follows from eq. (\ref{cons}).

To select the double poles in $k^{2}$ and $q^{2}$ we have to discard, in
eqs. (\ref{grR}) and (\ref{grNS}), those terms which are proportional to
$k^{2}$ or $q^{2}$ or $f(\tau)$ or $f(l^{\prime})$. In the $RR$ sector we
find, after multiplying by the $l\rightarrow \infty$ limit of $Z^{R+}$,
a result proportional to
\beq
\frac 1{k^2 q^2} \left[
h_{ij} k^i k^j - p \tanh (v_1 - v_2) h_{i1} k^i \right] 
\cosh (v_1 - v_2)\;. \label{74}
\eeq
This is in fact seen to correspond to the diagram, Fig. 2, where the
on-shell
outgoing graviton is coupled to the RR vector exchanged by the branes
through the minimal coupling  
\begin{equation}
\label{min}
{\cal L}_{Int} = h_{ij} T^{ij} \;.
\end{equation}
$T^{ij}$ is the symmetrized energy-momentum tensor of
the two RR vectors. The latter is given by 
\begin{equation}
T^{ij} = F^{i\alpha}_1 F^j_{2\alpha} - \frac 14 \eta^{ij} 
F^\alpha_{1\alpha} F^\beta_{2\beta}\;. 
\end{equation}
The second piece does not contribute upon contraction with the 
traceless polarisation tensor $h^{ij}$ of the graviton. The first 
part can be computed using the fields
\begin{equation}
A_1^\mu = \frac 1{k^2} J_1^\mu \;,\;\;
A_2^\mu = \frac 1{q^2} J_2^\mu \;.
\end{equation} 
The Feynman diagram is then found to give
\begin{eqnarray}
{\cal L}_{Int} = \frac 1{k^2 q^2} &&\left\{
h_{ij} k^i k^j \cosh (v_1 - v_2) 
- p h_{i1} k^i \sinh (v_1 - v_2) \right. \nonumber \\
&& \left. \;\; - k \cdot q h_{11} \sinh v_1 \sinh v_2 \right\} \;.
\end{eqnarray}
Since $k \cdot q = \frac 12 (k^2 + q^2)$, the last term does not
contribute to the double pole and we find an expression proportional to
eq. (\ref{74}).

In the NS-NS sector, there are two contributions, one coming from 
dilaton exchange and one from graviton exchange. 
For the dilatons, one has a minimal coupling of the form 
(\ref{min}). The symmetrized energy-momentum tensor of the two dilatons
is given by
\begin{equation}
T^{ij} \sim \partial^i \phi_1 \partial^j \phi_2 \;.
\end{equation}
Using the fields
\begin{equation}
\phi_1 = \frac 1{k^2} \;,\;\;
\phi_2 = \frac 1{q^2} \;,
\end{equation}
the interaction is found to be
\begin{equation}
{\cal L}_{Int} = \frac 1{k^2 q^2} h_{ij} k^i k^j \;.
\end{equation}
For the gravitons, the interaction can be deduced from the three 
gravitons vertex in the harmonic gauge \cite{grav}
\begin{eqnarray}
{\cal L}_{Int} = && \partial^\mu h^\alpha_\beta h^\nu_\alpha 
\partial_\nu h^\beta_\mu - \frac 12 h^\alpha_\beta \partial^\mu
h^\beta_\nu \partial_\mu h^\nu_\alpha + \frac 12 h^\alpha_\beta
\partial^\nu h^\mu_\alpha \partial_\mu h^\beta_\nu
+ \frac 12 h^\alpha_\beta \partial_\mu h^\beta_\alpha \partial^\mu h 
\nonumber \\ && + \frac 14 \partial_\mu h^\alpha_\beta \partial^\nu
h^\beta_\alpha h^\mu_\nu + \frac 12 \partial_\nu \partial^\mu h 
h^\beta_\mu h^\nu_\beta 
- \frac 14 h \partial_\nu h^\alpha_\beta \partial^\beta 
h^\nu_\alpha + \frac 18 h \partial^\beta h^\mu_\nu \partial_\beta
h^\nu_\mu \nonumber \\ && - \frac 18 h \partial^\mu h \partial_\mu h 
- \frac 14 h \partial^\nu \partial_\beta h h^\beta_\nu + \frac 12
h^\mu_\nu \partial^\alpha \partial_\beta h^\nu_\mu h^\beta_\alpha \;.
\end{eqnarray}
One has to choose in all possible ways one of the gravitons to be on-shell, 
with polarisation tensor satisfying ${\partial}_ih^{ij}=0$ and
$h^i_i=0$, and the other two to be the off-shell gravitons $h_1^{\mu \nu}$
and $h_2^{\mu \nu}$ coming from the two branes. Using the fields in the
harmonic gauge
\beq
h_1^{\mu \nu}=\frac{1}{k^2}(T_1^{\mu \nu}-\frac{1}{2}\eta^{\mu \nu}T_1)
\;,\;\;
h_2^{\mu \nu}=\frac{1}{q^2}(T_2^{\mu \nu}-\frac{1}{2}\eta^{\mu \nu}T_2)\;,
\eeq
and discarding terms containing $q^2$ or $k^2$, in the numerator, the
interaction is seen to be 
\beqa
{\cal L}_{Int}= \frac{1}{k^{2}q^{2}}&&\left\{-\frac{1}{4}h_{ij}k^{i}k^{j}
\cosh 2(v_{1}-v_{2})+\frac{p}{2}\sinh 2(v_{1}-v_{2})h_{1i}k^{i}\right.
\nn \\ &&\left.\;\,
-\frac{p^{2}}{2}h_{11}\sinh^{2}(v_{1}-v_{2})\right\}\;.
\eeqa

Now let us consider 
$({\cal M}^{NS+}_{grav}Z^{NS+}-{\cal M}^{NS-}_{grav}Z^{NS-})$ and look
for the double pole in $k^{2}$ and $q^{2}$. In the $l\rightarrow \infty$
limit we find an expression proportional to 
\beqa
&&\frac{1}{k^{2}q^{2}}\left\{-\frac{1}{4}h_{ij}k^{i}k^{j}
\cosh 2(v_{1}-v_{2})+\frac{p}{2}\sinh 2(v_{1}-v_{2})h_{1i}k^{i}-
\frac{p^{2}}{2}h_{11}\sinh^{2}(v_{1}-v_{2})\right\}\nonumber\\
&&+\frac{1}{k^{2}q^{2}}\left\{-\frac{1}{4}h_{ij}k^{i}k^{j}\sum_{a}\cos
2\pi z_{a}\right\}\;.
\eeqa
The first bracket matches the contribution from field theory where the
emitted graviton couples to the graviton exchanged between the branes
via the three graviton coupling. The second bracket matches the emission
of the graviton from scalar exchange, the factor $\sum_{a}cos2\pi z_{a}$
indicating that this possible scalar is related to the compactified
coordinates. In particular, for the D3-brane case of ref. \cite{hins}, the
invariant projection over the orbifold group gives 
$<\sum_{a}\cos 2\pi z_{a}=0>$, and there is no scalar emission
from the branes.

\section{Amplitudes for Other Massless Particles}

Concerning other massless particles, corresponding to other components of
the ten dimensional polarisation tensor $h_{AB}$ ($A,B=0,1,\cdots,9$), we
can restrict to the case where $A,B$ are space $i,j$ (transverse to
$\vec{p}$) or internal indices $a,b$. (For the internal indices we use a
complexified notation like $a=(4+i5),(6+i7),(8+i9)$ meaning $X^4+iX^5$, 
etc. and similarly for $a^*$. We do the same for the fermionic coordinates,
see ref. \cite{hins}.) The case where they are both space has been already
discussed. In the case where $A=i$ and $B=b$ the matrix element of the
vertex is zero for the branes we have considered and disregarding non zero
compactified momenta for the large distance limit. In the case where $A=a$ 
and $B=a^*$ we can have a non zero result for D0-branes compactified either 
on $T^6$ or on an orbifold, with the same Neumann or Dirichlet boundary 
conditions for both members of pairs of compactified coordinates (described 
by the boundary state of eq. (10) of ref. \cite{hins}). 
This is consistent with our previous work \cite{hins} where we
have seen that for these compactifications the branes are coupled
to spacetime scalars, which can then be emitted.
Technically, the non-zero result comes because there is no
term $h_{ij}k^ik^j$ in the amplitude eq. (\ref{dilaton}), and thus the
previously seen cancellation does not occur.

Finally in the case of orbifold compactification with mixed
Neumann-Dirichlet boundary conditions for the pairs of compactified
coordinates (corresponding to D3-branes described by the boundary state of
eqs (12), (14) of ref. \cite{hins}), the non zero matrix element occurs
for $A=a$ and $B=a$. But in this case the projection over the orbifold
invariant states multiply the vertex by $1+g_a^2+g_a^4=0$ (with
$g_a=exp(\pm i2\pi /3)$), thus there is no emission, consistently with the
analysis of ref. \cite{hins}.

Let us finally recall that we showed in Sec. IV that there is no emission
of the spacetime dilaton in all cases. By spacetime dilaton we mean the
massless scalar corresponding to the trace part of the four dimensional
polarisation tensor. The spacetime dilaton looks to be uncoupled to the
branes, consistently with an analysis appearing in ref. \cite{divec2}. There
can only occur, in some case, the emission of dilaton-like
scalars corresponding to the trace of the compactified (internal)
components of the polarisation tensor.

\begin{center}{\bf ACKNOWLEDGEMENTS}\end{center}
The authors would like to thank E. Gava and K.S. Narain for valuable
discussions. C.N. would like to thank the High Energy Group of ICTP for 
hospitality during the completion of this work which was done in the 
framework of the Associate Membership Programme of the ICTP. C.A.S.
would also like to thank J.F. Morales and M. Serone for useful
discussions. R.I.
and C.A.S. acknowledge partial support from EEC contract ERBFMRXCT96-0045.

\appendix

\section{Spacetime boundary state}

In this section we briefly recall the construction
of the spacetime part of the boundary state for a D0-brane.
Starting from the static case, in which the boundary conditions
are Neumann for the time direction and Dirichlet for the space
directions, the boost required
to get the dynamical case is easily implemented as an imaginary 
rotation. We will use the complex variable $z = \sigma + i \tau$, 
with $\sigma$ periodic and ranging from $0$ to $1$, and $\tau$
ranging from $0$ to $l$.

Starting with the bosonic coordinates, recall the mode expansion

\begin{eqnarray}
X^\mu(z) = &&\frac {X_o^\mu}2 -  \frac z 2 Q^\mu
+ \frac i{\sqrt{4\pi}} \sum_{n>0} \frac 1{\sqrt{n}} 
(a_n^{\mu} e^{2\pi n i z} - a_{n}^{\dagger \mu} e^{-2\pi n i z})\;, 
\nonumber \\
\bar X^\mu (\bar z)= &&\frac {X_o^\mu}2 +  \frac {\bar z} 2 Q^\mu 
+ \frac i{\sqrt{4\pi}} \sum_{n>0} \frac 1{\sqrt{n}} 
(\tilde a_n^{\mu} e^{-2\pi n i \bar z} - \tilde a_{n}^{\dagger \mu} 
e^{2\pi n i \bar z}) \;,
\end{eqnarray}
with the standard commutation relations $[a_m^\mu,a_n^{\dagger \nu}] 
=[\tilde a_m^\mu,\tilde a_n^{\dagger \nu}] = \eta^{\mu\nu} 
\delta_{mn}$, and $[X_o^\mu,Q^\nu]= i\eta^{\mu\nu}$.

The static boundary conditions for the oscillators are
\begin{equation}
(a^0_n+\tilde{a}^{\dagger 0}_{n})|B>^B_{osc}=0 \;,\;\; 
(a^i_n-\tilde{a}^{\dagger i}_{n})|B>^B_{osc}=0 \;.
\end{equation}
Pairing the $X^0,X^1$ coordinates in the light-cone combinations
$X^\pm=X^0 \pm X^1$, whose oscillators ($\alpha_n=a^{0}_{n}+a^{1}_{n}, 
\beta_n=a^{0}_{n}-a^{1}_{n}$) have 
as the only non-vanishing commutation relations $[\alpha_m, \beta_n^\dagger] =
[\beta_m, \alpha_n^\dagger]=-2\delta_{mn}$, the boundary conditions become
\begin{eqnarray}
&&(\alpha_n+\tilde{\beta}^\dagger_{n})|B>^b_{osc}=0\;,\;\;
(\beta_n+\tilde{\alpha}^\dagger_{n})|B>^b_{osc}=0 \;, \nonumber\\
&&(a^{T i}_n-\tilde{a}^{T \dagger i}_{n})|B>^b_{osc}=0 \;,
\end{eqnarray}
and the oscillator part of the bosonic boundary state is written as
\begin{equation}
|B>^b_{osc}=\exp {\sum_{n=1}^{\infty}\{\frac{1}{2}
(\alpha^\dagger_{n}\tilde{\alpha}^\dagger_{n}
+\beta^\dagger_{n}\tilde{\beta}^\dagger_{n})
+a^{T \dagger}_{n}\tilde{a}^{T \dagger}_{n}\}}|0> \;.
\end{equation}
For the zero modes, the boundary conditions are
\begin{equation}
Q^0 |B,Y>^b_o = 0 \;,\;\; (X^i_o - Y^i)|B,Y>^b_o = 0 \;,
\end{equation}
where $Y^i$ is the transverse position of the brane. These are solved 
taking
\begin{equation}
|B,Y>^b_o = \delta^{(3)}(X_o^i - Y^i)|0> 
= \int \frac {d^3 \vec q}{(2 \pi)^3} e^{-i \vec Y \cdot \vec q}|\vec q> \;.
\end{equation}

In the dynamical case, it is convenient to introduce the rapidity 
$v$ related to the velocity by $V=\tanh v$. 
The boundary conditions then become
in terms of the light-cone combinations 
\begin{eqnarray}
&&(e^{-v}\alpha_n+e^{v}\tilde{\beta}^\dagger_{n})|B,V>^b_{osc}=0\;,\;\;
(e^{v}\beta_n+e^{-v}\tilde{\alpha}^\dagger_{n})|B,V>^b_{osc}=0 \;,\nonumber \\
&&(a^{T i}_n-\tilde{a}^{T \dagger i}_{n})|B,V>^b_{osc}=0\;.
\end{eqnarray}
Thus, the oscillator part of the boosted bosonic boundary state is 
\begin{equation}
|B,V>^b_{osc}=\exp{\sum_{n=1}^{\infty}\{\frac{1}{2}
(e^{-2v}\alpha^\dagger_{n}\tilde{\alpha}^\dagger_{n}
+e^{2v}\beta^\dagger_{n}\tilde{\beta}^\dagger_{n})
+a^{T \dagger}_{n}\tilde{a}^{T \dagger}_{n}\}}|0>\;.
\end{equation}
For the zero modes, the new boundary conditions are
\begin{eqnarray}
&&(\cosh v Q^0 - \sinh v Q^1) |B,V,Y>^b_o = 0 \;, \nonumber\\
&&(\cosh v X^1_o - \sinh v X^0_o - Y^1)|B,V,Y>^b_o = 0 \;, \nonumber\\
&&(X^{T i}_o - Y^{T i})|B,V,Y>^b_o = 0 \;,
\end{eqnarray}
where $Y^i$ is the transverse position of the brane. These are solved 
taking
\beqa
|B,V,Y>^b_o &&= \delta(\cosh v X^1_o - \sinh v X^0_o - Y^1) 
\delta^{(2)}(X_o^{T i} - Y^{T i})|0> \nn \\ && 
= \int \frac {d^3 \vec q}{(2 \pi)^3} e^{-i \vec Y \cdot \vec q}|q^\mu>\;, 
\eeqa
where $q^\mu = (\sinh v q^1, \cosh v q^1, \vec q_T)=(\gamma V q^1,
\gamma q^1, \vec q_T)$

In a more formal way, the moving boundary state is obtained from the 
static one with a boost of opposite velocity \cite{divec}:
\begin{equation}
|B,V,Y> = e^{-iv J^{01}}|B,Y>\;,
\end{equation}
where
\begin{equation}
J^{\mu \nu} = X_o^\mu Q^\nu - X_o^\nu Q^\mu - i \sum_{n=1}^{\infty}
(a_n^{\dagger \mu} a_n^\nu - a_n^{\dagger \nu} a_n^\mu
+\tilde a_n^{\dagger \mu} \tilde a_n^\nu - \tilde a_n^{\dagger \nu} 
\tilde a_n^\mu)\;.
\end{equation}

Now consider the fermionic part. The mode expansions  are
\begin{eqnarray}
&&\psi^\mu(z)=\sum_{n> 0} (\psi^\mu_n e^{2\pi n i z}
+ \psi^{\dagger \mu}_n e^{- 2\pi n i z})\;,\nonumber \\
&&\bar \psi^\mu(\bar z)=\sum_{n> 0} (\tilde \psi^\mu_n 
e^{-2\pi n i \bar z} + \tilde \psi^{\dagger \mu}_n e^{ 2\pi n i \bar
z})\;,\label{fermod}
\end{eqnarray}
where the sums are over half integer or integer depending on whether we
have NSNS or RR fermions. In the RR sector there are also zero modes 
$\psi_o^\mu$ and $\tilde \psi_o^\mu $.
The anticommutation relation for the oscillators are the standard ones, 
$\{\psi_m^\mu,\psi_n^{\dagger \nu}\} = 
\{\bar \psi_m^\mu,\bar \psi_n^{\dagger \nu}\}=
\eta^{\mu \nu} \delta_{mn}$, in both the NSNS and RR sectors, whereas the
RR zero modes satisfy the Clifford algebra $\{\psi_o^\mu,\psi_o^\nu\}=
\{\tilde \psi_o^\mu, \tilde \psi_o^\nu\}= \eta^{\mu \nu}$.

The static boundary conditions, consistent with the mode expansion
(\ref{fermod}) are
\begin{equation}
(\psi^0_n+i\eta\tilde{\psi}^{\dagger 0}_{n})|B,\eta>^f_{osc}=0 \;,\;\;
(\psi^i_n-i\eta\tilde{\psi}^{\dagger i}_{n})|B,\eta>^f_{osc}=0\;,
\end{equation}
where $\eta = \pm 1$ has been introduced to deal with the GSO projection.

Pairing the $\psi^0,\psi^1$ fields in the light-cone combinations
$\psi^A=\psi^0 + \psi^1,\,\,\psi^B=\psi^0 - \psi^1$, whose oscillators
$\psi^A_n, \psi^B_n$ have 
as the only non-vanishing commutation relation $\{\psi^A_m,
\psi_n^{B\dagger}\} =
\{\psi^B_m, \psi_n^{A\dagger}\}=-2\delta_{mn}$, both in the NSNS and RR 
sectors, the boundary conditions become
\begin{eqnarray}
&&(\psi^A_n+i\eta\tilde{\psi}^{B\dagger}_{n})|B,\eta>^f_{osc}=0\;,\;\;
(\psi^{B}_n+i\eta\tilde{\psi}^{A\dagger}_{n})|B,\eta>^f_{osc}=0 \;,\nonumber\\
&&(\psi^{T }_n-i\eta\tilde{\psi}^{T \dagger }_{n})|B,\eta>^f_{osc}=0\;,
\end{eqnarray}
and the oscillator part of the fermionic boundary state is written as
\begin{equation}
|B,\eta>^f_{osc}=\exp{i\eta\sum_{n>0}^{\infty}\{\frac{1}{2}
(\psi^{A\dagger}_{n}\tilde{\psi}^{A\dagger}_{n}
+\psi^{B\dagger}_{n}\tilde{\psi}^{B\dagger}_{n})
-\psi^{T \dagger}_{n}\tilde{\psi}^{T \dagger}_{n}\}}|0>\;,
\end{equation}
with appropriate moding for each sector.
For the RR zero modes, the boundary conditions are
\begin{equation}
(\psi^0_o+i\eta\tilde{\psi}^{\dagger 0}_{o})|B,\eta>^{R}_{o}=0 \;,\;\;
(\psi^i_o-i\eta\tilde{\psi}^{\dagger i}_{o})|B,\eta>^{R}_{o}=0\;.
\end{equation}

It is convenient to define $a=(\gamma^{0}+\gamma^{1})/2$,
$a^{*}=(\gamma^{0}-\gamma^{1})/2$ and $b=(-i\gamma^{2}+\gamma^{3})/2$,
$b^{*}=(-i\gamma^{2}-\gamma^{3})/2$ such that
$\{a,a^{*}\}=\{b,b^{*}\}=1$, and similarly for $\tilde{a}$,$\tilde{b}$,
all other anticommutators being zero. The boundary conditions for the
zero modes can then be written as
\begin{eqnarray}
&&(a+i\eta\tilde{a}^{*})|B,\eta>^{R}_o=0 \;,\;\; 
(a^{*}+i\eta\tilde{a}^)|B,\eta>^{R}_o=0 \;,\nonumber\\
&&(b-i\eta\tilde{b})|B,\eta>^{R}_o=0\;,\;\;
(b^{*}-i\eta\tilde{b}^{*})|B,\eta>^{R}_o=0\;.
\end{eqnarray}
Defining a ``vacuum'' $|0>\otimes|\tilde{0}>$ by
$a|0>=b|0>=\tilde{a}|\tilde{0}>=\tilde{b}^{*}|\tilde{0}>=0$, we find the
zero mode stationary boundary state
\begin{equation}
|B,\eta>^{R}_{0}= \frac{1}{\sqrt{2}}e^{-i\eta(a^{*}\tilde{a}^{*}
-b^{*}\tilde{b})} |0>\otimes|\tilde{0}> \;.
\end{equation}
Notice that the boundary conditions imply that for $z=\bar z$, i.e. $\tau = 0$,
\begin{equation}
\psi^0(z)=-i\eta \bar \psi^0(\bar z)\;,\;\;
\psi^i(z)=i\eta \bar \psi^i(\bar z)\;.
\end{equation}

In the dynamical case, the boundary conditions become
in terms of the light-cone combinations $\psi^{A,B}$, 
\begin{eqnarray}
&&(e^{-v}\psi^A_n+i\eta
e^{v}\tilde{\psi}^{B\dagger}_{n})|B,\eta,V>^f_{osc}=0
\;,\;\; (e^{v}\psi^B_n+i\eta e^{-v}\tilde{\psi}^{A\dagger}_{n})
|B,\eta,V>^f_{osc}=0 \;,\nonumber \\
&&(\psi^{T }_n-i\eta\tilde{\psi}^{T \dagger }_{n})|B,\eta,V>^f_{osc}=0\;,
\end{eqnarray}
so that the oscillator part of the boosted fermionic boundary state is 
\begin{equation}
|B,\eta,V>^f_{osc}=\exp{i\eta\sum_{n>0}^{\infty}\{\frac{1}{2}
(e^{-2v}\psi^{A\dagger}_{n}\tilde{\psi}^{A\dagger}_{n}
+e^{2v}\psi^{B\dagger}_{n}\tilde{\psi}^{B\dagger}_{n})
-\psi^{T \dagger}_{n}\tilde{\psi}^{T \dagger}_{n}\}}|0>\;,
\end{equation} 
with appropriate moding for each sector. \newline
For the RR zero modes, the new boundary conditions are
\begin{eqnarray}
&&(e^{-v}a+i\eta e^{v}\tilde{a}^{*})|B,\eta>^{R}_o=0 \;,\;\;
(e^{v}a^{*}+i\eta e^{-v} \tilde{a}^)|B,\eta>^{R}_o=0 \;,\nonumber\\ 
&&(b-i\eta\tilde{b})|B,\eta>^{R}_o=0 \;,\;\;
(b^{*}-i\eta\tilde{b}^{*})|B,\eta>^{R}_o=0\;.
\end{eqnarray}
The boosted zero mode boundary state then becomes
\begin{equation}
|B,\eta,V>^{R}_{o}= \frac{1}{\sqrt{2}}e^{v}
e^{-i\eta(e^{-2v}a^{*}\tilde{a}^{*} -b^{*}\tilde{b})} |0>\otimes|\tilde{0}>\;.
\end{equation}
Notice furthermore that the new boundary conditions imply that for 
$z=\bar z$, i.e. $\tau = 0$,
\begin{eqnarray}
\label{fermcond}
&&\psi^0(z)=-i\eta(\cosh 2v \bar \psi^0(\bar z) 
- \sinh 2v \bar \psi^1(\bar z))\;,\nonumber \\
&&\psi^1(z)=i\eta((\cosh 2v \bar \psi^1(\bar z) 
- \sinh 2v \bar \psi^0(\bar z))) \;,\nonumber\\
&&\psi^{T }(z)=i\eta\bar \psi^{T }(\bar z)\;.
\end{eqnarray}

\section{Partition functions and Propagators}

In this section, we will use the boosted boundary states to compute
the uncompactified part of the partition functions and the correlation
functions on the cylinder. The contribution to the partition function of
the $(2,3)$ bosonic and fermionic coordinates cancels with the ghost
contributions (except that in the odd spin structure case the
$\beta-\gamma$ ghosts always contain the zero mode insertion). The net
effect of the velocity
is a twist. We shall define the modular parameter $q=e^{-2 \pi l}$.

For the bosonic field, we need only to consider the oscillator part, with
the Hamiltonian
\begin{equation}
H_{osc}=2\pi \sum_{n=1}^{\infty} n\left\{-{1\over 2} 
(\alpha^\dagger_{n} \beta_{n} + \beta^\dagger_{n} \alpha_n 
+\tilde \alpha^\dagger_{n} \tilde \beta_n +\tilde \beta^\dagger_{n} 
\tilde \alpha_n) + a^{T \dagger}_{n} a_{n}^T + 
\tilde a_{n}^{\dagger T} \tilde a_{n}^T) \right\}\;.
\end{equation}
Here $a^T$ includes all the transverse directions, both uncompactified and
compactified.
The uncompactified part of the partition function 
\begin{equation}
Z^b=<B,V_1|e^{-l H_{osc}}|B,V_2>^b_{osc}
\end{equation}
is then computed to be (taking into account the ghost contribution)
\begin{equation}
Z^b(unc)=\prod_{n=1}^{\infty}\frac 1{\left(1 - q^{2n} e^{-2(v_1 - v_2)} 
\right) \left(1 - q^{2n} e^{2(v_1 - v_2)} \right)}\;.\label{zbos}
\end{equation}
The complete partition function has been explicitly written in ref.
\cite{hins}, eq. (16) (for the toroidal compactification or in general for
D0-branes putting $z_a=0$) and in eq. (19) (for D0-branes on a $Z_3$
orbifold fixed point, twisted sector). 

The correlation functions, as defined in eq. (\ref{ave}) require a bit
more work. Define
\begin{eqnarray}
A_v(\tau,l)&\equiv &<X^0(z) \bar X^0(\bar z)>_{osc}=
<X^1(z) \bar X^1(\bar z)>_{osc} \;,\nonumber \\
A(\tau,l)\delta^{ij}&\equiv &<X^{Ti}(z) \bar X^{Tj}(\bar z)>_{osc}
\;, \nonumber \\
B_v(\tau,l)&\equiv& <X^0(z) \bar X^1(\bar z)>_{osc}=
<X^1(z) \bar X^0(\bar z)>_{osc}\;,\nonumber  \\
C_v(l)&\equiv &<X^0(z) X^0(z)>_{osc}= <\bar X^0(\bar z) \bar X^0(\bar
z)>_{osc}\nonumber \\ 
&=& -<X^1(z) X^1(z)>_{osc}
=-<\bar X^1(\bar z) \bar X^1(\bar z)>_{osc}\;,\nonumber  \\
-  C(l)\delta^{ij}&\equiv &<X^{Ti}(z) X^{Tj}(z)>_{osc}=
<\bar X^{Ti}(\bar z) \bar X^{Tj}(\bar z)>_{osc}\;,
\end{eqnarray}
with $A(\tau,l)= A_v(\tau,l)|_{v_1=v_2=0}$ and $C(\tau,l) = 
C_v(\tau,l)|_{v_1=v_2=0}$.
Doing the oscillator algebra, and using the formulae 
\beq
\sum_{k=0}^{\infty}x^k= \frac 1{1 - x}\;,\;\;
\sum_{k=1}^{\infty}\frac {x^k}k=-\ln (1 - x)\;, 
\eeq
we write the results as
\begin{eqnarray}
&&A_v = \frac 1{4 \pi} \sum_{n=0}^{\infty}
\left\{\cosh 2[(v_1 - v_2)n - v_2] \ln (1 - q^{2n} e^{-4\pi\tau}) \right. 
\nonumber \\ && \qquad \qquad \qquad + \left.
\cosh 2[(v_2 - v_1)n - v_1] \ln (1 - q^{2n} e^{-4\pi l^\prime})
\right\}\;,\nonumber \\
&&B_v = -\frac 1{4 \pi} \sum_{n=0}^{\infty}
\left\{\sinh 2[(v_1 - v_2)n - v_2] \ln (1 - q^{2n} e^{-4\pi\tau}) \right.
\nonumber \\ && \qquad \qquad \qquad \;\; + \left.
\sinh 2[(v_2 - v_1)n - v_1] \ln (1 - q^{2n} e^{-4\pi l^\prime})
\right\}\;,\nonumber \\
&&C_v = \frac 1{2 \pi} \sum_{n=1}^{\infty}
\cosh 2[(v_1 - v_2)n] \ln (1 - q^{2n})\;.
\end{eqnarray}
In the last expression, we have discarded a normal ordering constant that
will never contribute in the amplitude because of $p^2=0$.

The bosonic exponential correlation is given by
\begin{equation}
\label{exp}
<e^{i p \cdot X}>_{osc}=e^{-\frac 12 p_\mu p_\nu <(X + \bar X)_{osc}^\mu 
(X + \bar X)_{osc}^\nu>}=e^{-[(p_0^2 + p_1^2) A_v 
+ \vec{p}_T^2 (A + C_v - C) + 2 p_0 p_1 B_v]}
\end{equation}
and, using $p=p^0$ and $\cos \theta = \frac {p^1}p$, can be recast 
in the following form
\begin{eqnarray}
<e^{i p \cdot X}>_{osc} = && \prod_{n=1}^{\infty} 
\left[ 1 - q^{2n} \right]^{- \frac {p^2}\pi 
\sinh^2 [(v_1 - v_2)n] \sin^2 \theta} \nonumber \\
&&\times \prod_{n=0}^{\infty} 
\left[1 - q^{2n} e^{-4\pi\tau} \right]^{-\frac{p^2}{2\pi}
\cosh^2 [(v_1 - v_2)n - v_2] \left\{1 + \tanh [(v_1 - v_2)n - v_2] 
\cos \theta \right\}^2} \nonumber \\
&&\times \prod_{n=0}^{\infty} \left[1 - q^{2n} e^{-4\pi l^\prime} 
\right]^{-\frac{p^2}{2\pi}
\cosh^2 [(v_2 - v_1)n - v_1] \left\{1 + \tanh [(v_2 - v_1)n - v_1] 
\cos \theta \right\}^2}\;.\label{exp2}
\end{eqnarray}

Consider now correlations involving only one derivative, and define 
\begin{eqnarray}
\frac i2 K_v(\tau,l)&\equiv & <\partial X^0(z) \bar X^0(\bar z)>_{osc}=
<\partial X^1(z) \bar X^1(\bar z)>_{osc} \nonumber \\
& =&-<\bar \partial \bar X^0(\bar z) X^0(z)>_{osc}= 
-<\bar \partial \bar X^1(\bar z) X^1(z)>_{osc}\;,\nonumber \\
\frac i2 \delta^{ij} K(\tau,l)&\equiv &<\partial X^{Ti}(z) \bar
X^{Tj}(\bar z)>_{osc} =
-<\bar \partial \bar X^{Tj}(\bar z) X^{Ti}(z)>_{osc} \;,\nonumber\\
\frac i2 L_v(\tau,l)&\equiv &<\partial X^0(z) \bar X^1(\bar z)>_{osc}=
<\partial X^1(z) \bar X^0(\bar z)>_{osc} \nonumber \\
& =&-<\bar \partial \bar X^1(\bar z) X^0(z)>_{osc}=
-<\bar \partial \bar X^0(\bar z) X^1(z)>_{osc}  \;, \nonumber\\
\frac i2 W_v(l)&\equiv &<\partial X^0(z) X^1(z)>_{osc}=
-<\bar \partial \bar X^0(\bar z) \bar X^1(\bar z)>_{osc}\;, 
\end{eqnarray}
with $K(\tau,l)\equiv K_v(\tau,l)|_{v_1=v_2=0}$.
One obtains
\begin{eqnarray}
K_v &=& -\sum_{n=0}^{\infty}
\left\{\cosh 2[(v_1 - v_2)n - v_2] 
\frac {q^{2n} e^{-4\pi\tau}}{1 - q^{2n} e^{-4\pi\tau}} \right. 
\nonumber \\ &&  \qquad \quad \; - \left.
\cosh 2[(v_2 - v_1)n - v_1] 
\frac {q^{2n} e^{-4\pi l^\prime}}{1 - q^{2n} e^{-4\pi l^\prime}}
\right\}\;,\nonumber \\
L_v & =& \sum_{n=0}^{\infty}
\left\{\sinh 2[(v_1 - v_2)n - v_2]
\frac {q^{2n} e^{-4\pi\tau}}{1 - q^{2n} e^{-4\pi\tau}} \right.
\nonumber \\ &&  \qquad \; - \left.
\sinh 2[(v_2 - v_1)n - v_1] 
\frac {q^{2n} e^{-4\pi l^\prime}}{1 - q^{2n} e^{-4\pi l^\prime}} \right\} 
\;, \nonumber \\
W_v &=& -\frac {v_1 - v_2}{2 \pi l}
-2\sum_{n=1}^{\infty} \sinh 2[(v_1 - v_2)n] 
\frac {q^{2n}}{1 - q^{2n}}\;.
\end{eqnarray}

Turn now to the fermions, whose Hamiltonian is
\begin{equation}
H =2\pi \sum_{n>0}^{\infty} n\left\{-{1\over 2} 
(\psi^{A\dagger}_{n} \psi^B_{n} + \psi^{B\dagger}_{n} \psi^A_n 
+\tilde \psi^{A\dagger}_{n} \tilde \psi^B_n +\tilde \psi^{B\dagger}_{n} 
\tilde \psi^A_n) + \psi^{T \dagger}_{n} \psi_{n}^T + 
\tilde \psi_{n}^{\dagger T} \tilde \psi_{n}^T) \right\} 
\end{equation}
with appropriate moding in each sector.
The uncompactified part of the partition functions
\begin{eqnarray}
&&Z^{R}_\pm=<B,\pm ,V_1|e^{-l H}|B,+,V_2>^R\;,\nonumber \\
&&Z^{NS}_\pm=<B,\pm ,V_1|e^{-l H}|B,+,V_2>^{NS}\;,
\end{eqnarray}
are found to be (taking into account the ghost contribution)
\begin{eqnarray}
&&Z^{R}_+(unc)=4 \cosh (v_1 - v_2) \prod_{n=1}^{\infty}\left(1 + q^{2n} 
e^{-2(v_1 - v_2)} \right) \left(1 + q^{2n} e^{2(v_1 - v_2)}
\right)\;,\nonumber \\
&&Z^{R}_-(unc)=4 \sinh (v_1 - v_2) \prod_{n=1}^{\infty}\left(1 - q^{2n} 
e^{-2(v_1 - v_2)} \right) \left(1 - q^{2n} e^{2(v_1 - v_2)}
\right)\;,\nonumber \\
&&Z^{NS}_\pm(unc)=\prod_{n=1}^{\infty} \left(1 \pm q^{2n - 1} 
e^{-2(v_1 - v_2)} \right) \left(1 \pm q^{2n - 1} e^{2(v_1 - v_2)}
\right)\;.\label{parfunx}
\end{eqnarray}

The complete partition functions for the even spin structures have been
written in ref. \cite{hins} eqs. (28) (toroidal or generic orbifold case)
and (32) (D0-branes on a $Z_3$ orbifold fixed point, twisted sector). The
odd spin structure case has already been separately discussed in Sec. I. 

Let us discuss the fermionic correlation functions.
In order to treat the odd spin structure case, we have to make some 
preliminary observation. Referring to our definition eq. (\ref{aveodd}), 
we have to consider two cases, when ${\cal O}$ is quadratic in the fermion
operators and when it is quartic. In the quadratic case we define 
$<{\cal O}>^{odd}\equiv \ll{\cal O}\gg^{odd}$, and thus the only non
vanishing correlators are
\beq
\frac{1}{2}\epsilon^{ij}=<\psi^{Ti}(z)\bar \psi^{Tj}(\bar z)>^{odd}
= -i<\psi^{Ti}(z)\psi^{Tj}(z)>^{odd}
= i<\bar \psi^{Ti}(\bar z)\bar \psi^{Tj}(\bar z)>^{odd}\;,
\label{star}
\eeq 
with $\epsilon^{ij}=\epsilon^{1ij}$.
For the quartic case, the only non vanishing result is when ${\cal O}$ contains
the zero modes in the two transverse directions and each only once.
Thus, ${\cal O}$ can be written as a product of a quadratic term, say
$\psi\psi$ times the 2-3 zero modes (which can be either both left or
right or mixed):
${\cal O} = \psi\psi \cdot (0-modes)^{2,3}$.
We then define the odd propagator $<\psi\psi>^{odd}$ by
\beq
<\psi\psi>^{odd} <(0-modes)^{2,3}> \equiv\ll\psi\psi\circ
(0-modes)^{2,3}\gg^{odd}\;,\label{starodd}
\eeq
where $<(0-modes)^{2,3}>$ is given by eq. (\ref{star}) and $\ll \cdots\gg$
by eq. (\ref{aveodd}).

We thus have (here we always put $\eta_1=+1$), in all cases for $s=even$
and in the quartic case for $s=odd$,
\begin{eqnarray}
i F^s_v(\tau,l)&\equiv &<\psi^0(z) \bar \psi^0(\bar z)>^s=
<\psi^1(z) \bar \psi^1(\bar z)>^s \;, \nonumber \\
i\delta^{ij} F^{s(even)}(\tau,l)&\equiv&<\psi^{Ti}(z)\bar \psi^{Tj}(\bar
z)>^{s\,\,even} \;, \nonumber \\
i\delta^{ij} F^{odd}(\tau,l)&=&<\psi^{Ti}(z)\bar \psi^{Tj}(\bar
z)>^{odd} \;, \nonumber \\
i G^s_v(\tau,l)&\equiv &<\psi^0(z) \bar \psi^1(\bar z)>^s=
<\psi^1(z) \bar \psi^0(\bar z)>^s \;, \nonumber \\
U^s_v(l)&\equiv &<\psi^0(z) \psi^1(z)>^s=
<\bar \psi^0(\bar z) \bar \psi^1(\bar z)>^s \;, \nonumber  \\
0 &= &<\psi^{Ti}(z) \psi^{Tj}(z)>^{even}=
-<\bar \psi^{Ti}(\bar z) \bar \psi^{Tj}(\bar z)>^{even}\;, 
\label{star2}
\end{eqnarray}
with $F^s(\tau,l) = F^s_v(\tau,l)|_{v_1=v_2=0}$ both in the
even and the odd cases. We can use Wick's theorem evaluating matrix elements
by using the propagators defined in eq. (\ref{star2}) 
(which for the odd spin structure refers to
the four fermions case) and eq. (\ref{star}) (two fermions case).
Notice that we have also
$<\psi^{1}\psi^{Ti}> = <\bar{\psi}^{1}\bar{\psi}^{Ti}> =
<\psi^{1}\bar{\psi}^{Ti}> = 0$.

In the NSNS$\pm$ sectors only fermionic oscillator modes appear, whereas
in the RR$\pm$ sectors we have also the fermionic zero modes. Their
contributions are
\beqa
&&F^{oR+}_v = -\frac 12 \frac {\cosh (v_1 + v_2)}{\cosh (v_1 - v_2)}
\;,\;\; F^{oR-}_v = 
-\frac 12 \frac {\sinh (v_1 + v_2)}{\sinh (v_1 - v_2)}\;, \nonumber \\
&&G^{oR+}_v = -\frac 12 \frac {\sinh (v_1 + v_2)}{\cosh (v_1 - v_2)}
\;,\;\; G^{oR-}_v = 
-\frac 12 \frac {\cosh (v_1 + v_2)}{\sinh (v_1 - v_2)} \;, \nonumber\\
&&U^{oR+}_v = \frac 12 \tanh (v_1 - v_2) \;,\;\;
U^{oR-}_v = \frac 12 \coth (v_1 - v_2)\;.
\eeqa
The full correlators are then obtained as
\beq
F^{R\pm}_v = F^{o R\pm}_v + \tilde F^{R\pm}_v \;,\;\;
G^{R\pm}_v = G^{o R\pm}_v + \tilde G^{R\pm}_v \;,\;\;
U^{R\pm}_v = U^{o R\pm}_v + \tilde U^{R\pm}_v \;,
\eeq
where
\begin{eqnarray}
&&\tilde F^{R\pm}_v = -\sum_{n=0}^{\infty} (\mp)^n
\left\{\cosh 2[(v_1 - v_2)n - v_2] 
\frac {q^{2n} e^{-4\pi\tau}}{1 - q^{2n} e^{-4\pi\tau}} \right. 
\nonumber \\ && \qquad \qquad \qquad \qquad \quad \pm \left.
\cosh 2[(v_2 - v_1)n - v_1] 
\frac {q^{2n} e^{-4\pi l^\prime}}{1 - q^{2n} e^{-4\pi l^\prime}} 
\right\} \;,\nonumber\\
&&\tilde G^{R\pm}_v = \sum_{n=0}^{\infty} (\mp)^n
\left\{\sinh 2[(v_1 - v_2)n - v_2] 
\frac {q^{2n} e^{-4\pi\tau}}{1 - q^{2n} e^{-4\pi\tau}} \right. 
\nonumber \\ && \qquad \qquad \qquad \qquad \pm \left.
\sinh 2[(v_2 - v_1)n - v_1] 
\frac {q^{2n} e^{-4\pi l^\prime}}{1 - q^{2n} e^{-4\pi l^\prime}} 
\right\} \;, \nonumber\\
&&\tilde U^{R\pm}_v = -\frac {(v_1 - v_2)}{2\pi l}
-2\sum_{n=1}^{\infty} (\mp)^n
\sinh 2[(v_1 - v_2)n] \frac {q^{2n}}{1 - q^{2n}}\;,
\end{eqnarray}
in the RR sector and
\begin{eqnarray}
&&F^{NS\pm}_v = -\sum_{n=0}^{\infty} (\mp)^n
\left\{\cosh 2[(v_1 - v_2)n - v_2] 
\frac {q^{n} e^{-2\pi\tau}}{1 - q^{2n} e^{-4\pi\tau}} \right. 
\nonumber \\ && \qquad \qquad \qquad \qquad \quad \pm \left.
\cosh 2[(v_2 - v_1)n - v_1] 
\frac {q^{n} e^{-2\pi l^\prime}}{1 - q^{2n} e^{-4\pi l^\prime}} 
\right\} \;, \nonumber\\
&&G^{NS\pm}_v = \sum_{n=0}^{\infty} (\mp)^n
\left\{\sinh 2[(v_1 - v_2)n - v_2] 
\frac {q^{n} e^{-2\pi\tau}}{1 - q^{2n} e^{-4\pi\tau}} \right. 
\nonumber \\ && \qquad \qquad \qquad \qquad \pm \left.
\sinh 2[(v_2 - v_1)n - v_1] 
\frac {q^{n} e^{-2\pi l^\prime}}{1 - q^{2n} e^{-4\pi l^\prime}} 
\right\} \;, \nonumber\\
&&U^{NS\pm}_v = -\frac {(v_1 - v_2)}{2\pi l}
-2\sum_{n=1}^{\infty} (\mp)^n
\sinh 2[(v_1 - v_2)n] \frac {q^{n}}{1 - q^{2n}}\;,
\end{eqnarray}
in the NSNS sector. The equal-point correlators $U^{R\pm}$ and $U^{NS\pm}$
can be deduced from the other correlators by using the eq. (\ref{fermcond}) 
to reflect left and right movers at the boundaries.

Notice that world sheet supersymmetry is enforced between the bosons and
the RR odd spin structure fermions. Since $K_v = \tilde F^{R-}_v$, 
$L_v = \tilde G^{R-}_v$ and $W_v = \tilde F^{R-}_v$, 
we explicitly check the relations
\begin{eqnarray}
&&<\partial X^\mu(z) \bar X^\nu(\bar z)>_{osc} = 
\frac 12 <\psi^\mu(z) \bar \psi^\nu (\bar z)>^{R-}_{osc}\;, \nonumber \\
&&<\partial X^\mu(z) X^\nu(z)>_{osc} = 
\frac i2 <\psi^\mu(z) \psi^\nu (z)>^{R-}_{osc} \;, \nonumber\\
&&<\bar \partial \bar X^\mu(\bar z) X^\nu(z)>_{osc} = 
\frac \eta 2 <\bar \psi^\mu(\bar z) \psi^\nu (z)>^{R-}_{osc}\;, \nonumber \\
&&<\bar \partial \bar X^\mu(z) \bar X^\nu(\bar z)>_{osc} = 
-\frac i2 <\bar \psi^\mu(\bar z) \bar \psi^\nu (\bar z)>^{R-}_{osc}\;.
\label{sup}
\end{eqnarray} 

The periodicities of the fermionic propagators in the four spin structures,
which should follow from an involution from the torus to the cylinder,
can be seen considering the light-cone combinations $\psi^\pm = \psi^0 
\pm \psi^1$ and in particular their propagators $<\psi^\pm (z) 
\bar \psi^\pm (\bar z)>_s =  P_{v(\pm)}^s$, which are given by
\begin{equation}
P_{v(\pm)}^s = \frac i2 (F_v^s \pm G_v^s)\;.
\end{equation}
One can then explicitly check the transformation around the two cycles of 
the covering torus, which has modulus $\nu=2il$, 
$w \rightarrow w + m + \nu n$ with $w=z-\bar z=2i\tau$, that is 
$\tau \rightarrow \tau - \frac i2 m + n l$, getting
\begin{eqnarray}
&&P_{v (\pm)}^{R+}(\tau - \frac i2 m + n l, l) = 
e^{i \pi n \pm 2n(v_1 - v_2)} P_{v (\pm)}^{R+}(\tau, l) \;, \nonumber\\
&&P_{v (\pm)}^{R-}(\tau - \frac i2 m + n l, l) = 
e^{\pm 2n(v_1 - v_2)} P_{v (\pm)}^{R-}(\tau, l) \;, \nonumber\\
&&P_{v (\pm)}^{NS+}(\tau - \frac i2 m + n l, l) = 
e^{i \pi m + i \pi n \pm 2n(v_1 - v_2)} P_{v (\pm)}^{NS+}(\tau,l)
\nonumber \;, \\
&&P_{v (\pm)}^{NS-}(\tau - \frac i2 m + n l, l) = 
e^{i \pi m \pm 2n(v_1 - v_2)} P_{v (\pm)}^{NS-}(\tau, l)\;.
\end{eqnarray}
These transformation rules for $m=0$  correspond to the boundary conditions at the two ends of the
cylinder for the $\psi^\pm$ which are
\begin{eqnarray} 
&&\psi^\pm(z) |_{\tau=0}  = -i e^{\pm 2 v_2} 
\bar \psi^\mp (\bar z) |_{\tau=0}\;, \nonumber \\
&&\psi^\pm(z) |_{\tau=l} = -ie^{\pm 2 v_1} 
\bar \psi^\mp (\bar z) |_{\tau=l}\;.
\end{eqnarray}
The local behavior of these functions for $\tau \rightarrow 0$ is found 
to be
\begin{equation}
P_{v(\pm)}^s (\tau, l) \rightarrow \frac 1{8\pi i \tau} e^{\pm 2v_2}\;.
\end{equation}

It is convenient to rescale the fermions according to $\psi^\pm 
\rightarrow \hat \psi^\pm = e^{\mp v_2} \psi^\pm$, so that their 
propagators are $ \hat P_{v(\pm)}^s = e^{\mp 2 v_2} P_{v(\pm)}^s$. The
monodromy properties do not change, but the boundary conditions now become
\begin{eqnarray} 
&&\hat \psi^\pm(z) = -i \hat {\bar \psi}^\mp (\bar z)
\;,\;\; z = \bar z\;,\nonumber \\
&&\hat \psi^\pm(z) = -i e^{\pm 2 (v_1 - v_2)} \hat 
{\bar \psi}^\mp (\bar z)
\;,\;\; z = \bar z + \nu\;,
\end{eqnarray}
and the local behavior for $\tau \rightarrow 0$ simplifies to the 
conventional one
\begin{equation}
\hat P_{v(\pm)}^s (\tau, l) \rightarrow \frac 1{4\pi w}\;.
\end{equation}

It has become now clear how to do the twisted involution to pass from the 
covering torus to the cylinder: the twisted boundary conditions on the
cylinder are obtained from a non-trivial phase transformation around the
long cycle of the torus with imaginary angle 
$\epsilon = \frac {v_1 - v_2}\pi$. Actually,
the monodromy properties of the functions $\hat P_{v(\pm)}^s$, together with 
their local behavior, imply them to be combinations of 
twisted ${\cal \theta}$-functions, with argument $w=2i\tau$, 
modulus $\nu=2il$ and imaginary twist $\epsilon = \frac {v_1 - v_2}\pi$. 
In fact, one can check that
\begin{eqnarray}
\hat P_{v (\pm)}^{R+}(w, \nu) &=& \frac 1{4\pi}
\frac {\theta_2 (w \pm i \epsilon | \nu) \theta_1^\prime (0 | \nu)}
{\theta_1 (w | \nu) \theta_2 (\pm i \epsilon|\nu)}\;,\nonumber \\
\hat P_{v (\pm)}^{NS+}(w, \nu) &=& \frac 1{4\pi}
\frac {\theta_3 (w \pm i \epsilon | \nu) \theta_1^\prime (0 | \nu)}
{\theta_1 (w | \nu) \theta_3 (\pm i \epsilon|\nu)} \;,\nonumber\\
\hat P_{v (\pm)}^{NS-}(w, \nu) &=& \frac 1{4\pi}
\frac {\theta_4 (w \pm i \epsilon | \nu) \theta_1^\prime (0 | \nu)}
{\theta_1 (w | \nu) \theta_4 (\pm i \epsilon|\nu)} \;.
\end{eqnarray}

In order to study the amplitudes in the large distance limit, we will need
the $l \rightarrow +\infty$ asymptotics of the correlations.
For the bosonic exponential one gets
\begin{equation}
<e^{i p \cdot X}>_{osc} \;\rightarrow 
\left[1 - e^{-4\pi \tau} \right]^{-\frac {p_B^{(2)2}}{2\pi}}
\left[1 - e^{-4\pi l^\prime} \right]^{-\frac {p_B^{(1)2}}{2\pi}}\;,
\end{equation}
whereas the fermionic propagators in the four spin structures, in this
limit, reduce to
\begin{eqnarray}
\tilde F^{R\pm}_v \rightarrow && -\cosh 2v_2 
\frac {e^{-4\pi\tau}}{1 - e^{-4\pi\tau}} 
\mp \cosh 2v_1 \frac {e^{-4\pi l^\prime}}{1 - e^{-4\pi l^\prime}}\;,
\nonumber \\
\tilde G^{R\pm}_v \rightarrow && -\sinh 2v_2 
\frac {e^{-4\pi\tau}}{1 - e^{-4\pi\tau}} 
\mp \sinh 2v_1 \frac {e^{-4\pi l^\prime}}{1 - e^{-4\pi l^\prime}}\;,
\nonumber \\
\tilde U^{R\pm}_v \rightarrow && -\frac {(v_1 - v_2)}{2\pi l} 
\pm 2 \sinh 2(v_1 - v_2) e^{-4 \pi l} \;,
\end{eqnarray}
and
\begin{eqnarray}
F^{NS\pm}_v \rightarrow && -\cosh 2v_2 
\frac {e^{-2\pi\tau}}{1 - e^{-4\pi\tau}} 
\mp \cosh 2v_1 \frac {e^{-2\pi l^\prime}}{1 - e^{-4\pi
l^\prime}}\nn\\
&&+e^{-2\pi l}(\pm\cosh 2(v_1-2v_2)e^{-2\pi \tau}+\cosh
2(v_2-2v_1)e^{-2\pi l^{\prime}})\;,\nonumber\\ 
G^{NS\pm}_v \rightarrow && -\sinh 2v_2 \frac {e^{-2\pi\tau}}{1 -
e^{-4\pi\tau}} 
\mp \sinh 2v_1 \frac {e^{-2\pi l^\prime}}{1 - e^{-4\pi l^\prime}}\nn \\
&&+e^{-2\pi l}(\mp\sinh 2(v_1-2v_2)e^{-2\pi\tau}-\sinh 2(v_2-2v_1)
e^{2\pi l^{\prime}}) \;, \nonumber \\
U^{NS\pm}_v \rightarrow && -\frac {(v_1 - v_2)}{2\pi l}
\pm 2 \sinh 2(v_1 - v_2) e^{-2 \pi l}\;.
\end{eqnarray}

\end{document}